\documentclass[11pt]{article}
\usepackage{amssymb,amsmath}
\usepackage{bm} 
\usepackage{booktabs} 
\usepackage{array}
\usepackage{latexsym}
\usepackage{graphicx}
\usepackage{color}
\usepackage{datetime}
\usepackage[nosort]{cite}
\usepackage{verbatim}
\usepackage{enumerate}
\usepackage{chngpage} 
\usepackage{mathrsfs}
\usepackage{euscript}
\usepackage{psfrag}

\usepackage{datetime}

\usepackage[nosort]{cite}
\usepackage{chngpage} 
\usepackage{setspace}
\usepackage{tensor}

\usepackage{tensor}

\usepackage{mciteplus}

\usepackage[colorlinks=true,      linkcolor=blue,      urlcolor=blue,      
            filecolor=blue,      citecolor=blue,       pdfstartview=FitH,     
						pdfpagemode=UseNone,      bookmarksopen=true]{hyperref}  
\usepackage[all]{hypcap}

\definecolor{cardinal}{rgb}{0.6,0,0}
\definecolor{darkgreen}{rgb}{0,0.4,0}
\definecolor{golden}{rgb}{0.92, 0.7, 0}
\definecolor{midnight}{rgb}{0, 0, 0.5}
\definecolor{darkblue}{rgb}{0, 0, 0.7}
\definecolor{purple}{rgb}{0.5, 0, 0.5}

\newcommand{\AdS}{\mathrm{AdS}}
\newcommand{\vol}{\operatorname{vol}}
\newcommand{\tr}{\operatorname{tr}}

\newcommand{\dd}{\mathrm{d}}
\newcommand{\Lu}{\mathcal L_u}
\newcommand{\Lv}{\mathcal L_v}

\newcommand{\R}{\mathbb R}
\newcommand{\lp}{\ell_p}

\def\d{{d}}

\numberwithin{equation}{section}

\begin{document}

\phantom{AAA}
\vspace{-10mm}

\begin{flushright}

\end{flushright}

\vspace{1.9cm}

\begin{center}

{\huge {\bf   A New Holographic Duality for Conformal Mass Deformations}}\\
{\huge {\bf \vspace*{.25cm}  }}

\vspace{1cm}

{\large{\bf {Rapha\"el Dulac$^{1}$}}}
 
\vspace{1cm}

$^1$Institut de Physique Th\'eorique, \\
Universit\'e Paris Saclay, CEA, CNRS,\\
Orme des Merisiers, Gif sur Yvette, 91191 CEDEX, France \\[12pt]

\vspace{10mm} 
{\footnotesize\upshape\ttfamily raphael.dulac @ ipht.fr} \\

\vspace{2.2cm}
 
\textsc{Abstract}
 
\end{center}

\vspace{-1.2mm}
\noindent

We construct a space-dependent conformal mass deformation of ABJM on the half-space $z>0$ and show that the vacua of this theory are dual to a large class of half-BPS bubbling solutions of the form $\AdS_3\times S^3\times S^3\times\Sigma$, with superalgebra $D(2,1|\gamma)^{\oplus2}$, parametrized by $\gamma<0$. The conformal symmetry of a plane is preserved provided that the mass deformation scales as $m(z)=\frac{c}{z}$. We give the first field-theory explanation of the exceptional superalgebra parameter $\gamma$ in M2-brane field theory in terms of the coefficient of the inverse-distance mass deformation of ABJM: $\gamma=\frac{1-2c}{1+2c}$. We demonstrate this using the methods of Polchinski and Strassler, which we first apply to the Janus interfaces of $\mathcal{N}=4$ super Yang-Mills and ABJM to relate localized mass sources to scalar matching conditions.

\thispagestyle{empty}
\newpage


\tableofcontents


\section{Introduction}


Adding a constant mass term to a CFT introduces a scale and therefore breaks conformal invariance. A position-dependent mass, however, can preserve the conformal symmetry of a planar interface or boundary, provided its spatial dependence compensates for its scaling dimension. Since the mass has dimension one, two natural profiles are available: a delta function supported on the interface, $m(\tilde z)\propto \delta(\tilde z)$, as in Janus interfaces, and an inverse-distance profile $m(\tilde{z})=\frac{c}{\tilde{z}}$ on the half-space $\tilde z>0$. Here $\tilde{z}$ is the coordinate normal to the interface and $c$ is a dimensionless constant. The latter is an example of the conformal mass deformations studied in \cite{Herzog:2019bom}.

The holographic study of mass deformations was developed by Polchinski and Strassler for $\mathcal{N}=1^{*}$ super Yang-Mills \cite{Polchinski:2000uf}. An analogous analysis for the M2-brane field theory, which will be important for our study, was carried out in \cite{Bena:2000zb}. In these constructions, the fermion masses can be read from non-normalizable flux perturbations in supergravity, while probe brane reveal the scalar interactions and brane polarization. Fully backreacted solutions are generally difficult to obtain. An important exception is provided by the LLM geometries \cite{Bena:2004jw,Lin:2004nb}, which include the gravity duals of vacua of mass-deformed ABJM theory \cite{Aharony:2008gk,Aharony:2008ug,Kim:2010mr,Cheon:2011gv}. These examples concretely show how nontrivial vacuum data can be encoded in the topology and fluxes of a smooth geometry. In this paper, we apply the methods of Polchinski and Strassler to shed new light on Janus interfaces, and find the holographic dual of a large class of solutions of \cite{Bachas:2013vza}.

Position-dependent deformations preserving planar conformal symmetry provide another setting in which exact supergravity solutions are available. Their dual geometries contain an AdS factor associated with the conformal group of the interface. When sufficient supersymmetry and R-symmetry are preserved, the supergravity equations admit an integrable structure. D'Hoker, Estes, Gutperle and collaborators used this structure to construct broad families of solutions in type IIB supergravity \cite{DHoker:2006vfr,DHoker:2007hhe,DHoker:2007mci,DHoker:2007zhm} and M-theory \cite{DHoker:2008lup,DHoker:2008rje,DHoker:2008wvd}. Note that all  these families of supergravity solutions describe a variety of configurations, including interfaces, boundary conditions and Wilson loop operators of $\mathcal{N}=4$ SYM, ABJM and the M5-brane theory, they are not all dual to mass deformations.

The brane coordinates are central to this approach. Recent work has related broad classes of these $\AdS$ geometries to near-brane limits of intersecting brane systems preserving eight supercharges \cite{Bena:2023rzm,Bena:2024dre,Bena:2025hxt,Bena:2026vlv,Bena2026toappear}. These coordinates make explicit the underlying brane intersection structure, and make probe calculations particularly direct. We will extensively use these coordinates to understand the conformal mass deformations.

Our first examples of conformal mass deformations are the supersymmetric Janus interfaces of $\mathcal{N}=4$ super Yang-Mills and ABJM. In the type IIB construction, the jump in the Yang-Mills coupling is accompanied by a localized fermion bilinear and its supersymmetric scalar completion \cite{DHoker:2006qeo,Gaiotto:2008sd}. Related mass sources appear in M-theory Janus constructions \cite{Bobev:2013yra,Arav:2018njv}. We express these geometries in coordinates adapted to their underlying brane intersections and apply the methods of Polchinski and Strassler. We understand the localized fermion mass from the asymptotic flux , while the probe action determines the normalization and matching conditions of the transverse scalars. This analysis also explains how a floating brane can coexist with nontrivial mass terms in the dual field theory.

Our main application concerns the bubbling M-theory solutions with superalgebra $D(2,1|\gamma)\oplus D(2,1|\gamma)$ studied in \cite{Bachas:2013vza}. We focus on the $\gamma<0$ solutions, with an asymptotic $\AdS_4\times S^7$ region. We propose that these solutions describe conformal mass deformations of ABJM, obtained by specializing the spatially dependent supersymmetric construction of Kim and Kwon \cite{Kim:2018qle,Kim:2019kns} to:

\begin{equation}
    m(\tilde{z})=\frac{c}{\tilde{z}}\,,.
\end{equation}

The asymptotic four-form flux and the quadratic M2-brane probe action provide independent tests of the source dictionary. We also perform a field theory computation to argue for the enhancement of the field theory superalgebra to $D(2,1|\gamma)$. They identify the dimensionless mass coefficient $c$ with the supergravity parameter $\gamma$ through:
\begin{equation}
    \gamma=\frac{1-2c}{1+2c}\,,.
\end{equation}
In particular, the branch considered here corresponds to $|c|>\frac{1}{2}$.
The same deformation admits noncommuting supersymmetric scalar profiles. After extracting their prescribed dependence on $\tilde{z}$, their algebraic equations reduce to the GRVV vacuum equations\cite{Gomis:2008vc}. This provides a connection between matrix representations in the field theory and the discrete data of the bubbling geometries. We distinguish this classical correspondence from the additional quantum input needed to identify the allowed occupation numbers. 

The paper is organized as follows. Section \ref{sec:Brane intersections} introduces the brane intersections and their adapted coordinates. Sections \ref{sec:Janus interface in type IIB} and \ref{sec:Janus interface in Type Mtheory} apply the probe and flux analysis to Janus interfaces in type IIB supergravity and M-theory. Section \ref{sec:conformalMassdef} develops the conformal mass deformation of ABJM and argue for the new holographic duality. Readers who are only interested in the proposed holographic duality can skip \ref{sec:Janus interface in type IIB} and \ref{sec:Janus interface in Type Mtheory} .

The paper is organized as follows. Section\ref{sec:Brane intersections} introduces the brane intersections and their adapted coordinates. Sections \ref{sec:Janus interface in type IIB} and \ref{sec:Janus interface in Type Mtheory} apply the probe and flux analysis to the Janus interfaces of type IIB supergravity and M-theory. Section \ref{sec:conformalMassdef} develops the conformal mass deformation of ABJM and argues for the proposed holographic duality. Readers interested only in this duality can skip Section \ref{sec:Janus interface in type IIB} and the Janus analysis of Sections \ref{sec:JanusExpM} and \ref{sec:JanusProbepM}. Section \ref{sec:conformalMassdef} only uses the inhomogeneous mass deformation of ABJM and the M-theory solutions reviewed in the first two subsections of Section \ref{sec:Janus interface in Type Mtheory}.


\section{Preliminaries on brane intersections in brane coordinates}
\label{sec:Brane intersections}

\subsection{M2-M5-M5' intersection in M-theory}


Following \cite{Lunin:2007ab,Bena:2023rzm,Bena:2024dre}, we describe
the $\frac{1}{4}$-BPS M2-M5-M5$'$ system in coordinates adapted to
the supersymmetry projectors of the corresponding flat-brane
intersection. All three stacks share the directions $(x^0,x^1)$.
The M2-branes also extend along $z$, while the M5- and M5$'$-branes
span two orthogonal four-planes, parametrized by $\vec u$ and
$\vec v$, respectively:
\begin{center}
\begin{tabular}{c|cccc}
 & $x^{0,1}$ & $z$ & $\mathbb{R}_u^4$ & $\mathbb{R}_v^4$\\\hline
 M2 & $\times$ & $\times$ & &\\
 M5 & $\times$ & & $\times$ &\\
 M5$'$ & $\times$ & & & $\times$
\end{tabular}
\end{center}
Imposing rotational symmetries in the two $\mathbb{R}^4$ planes
and defining $u=|\vec u|$ and $v=|\vec v|$, the metric takes the form:
\begin{equation}
\begin{aligned}
ds_{11}^2 ~=~ e^{2A_0}\,\bigg[
&-(dx^0)^2+(dx^1)^2
+e^{-3A_0}(-\partial_z w)^{-\frac12}
 \left(du^2+u^2d\Omega_{u,3}^2\right)\\
&+e^{-3A_0}(-\partial_z w)^{\frac12}
 \left(dv^2+v^2d\Omega_{v,3}^2\right)
+(-\partial_z w)
 \left(dz+\frac{\partial_u w}{\partial_z w}\,du\right)^2
\bigg]\,,
\end{aligned}
\label{eq:11metric}
\end{equation}
where $A_0$ and $w$ depend on $(u,v,z)$, and $d\Omega_{u,3}^2$
and $d\Omega_{v,3}^2$ are the metrics on unit three-spheres.

The coordinate $z$ is fibered over the radial direction $u$.
This introduces an apparent asymmetry between the two four-planes,
which is a consequence of the coordinate choice. Locally, one may
instead use $w$ as a coordinate and regard $z$ as a function of
$(u,v,w)$. In this description, the roles of the two four-planes
are exchanged. A convenient orthonormal set of vielbein is:
\begin{equation}
e^0 =  e^{A_0}\, dx^0 \,, \qquad e^1=  e^{A_0}\, dx^1 \,,  \qquad e^2 =   e^{A_0}\,  (-\partial_z w )^{\frac{1}{2}} \, \left( dz +(\partial_z w )^{-1}\,  \big (\partial_u w \big)  d u \right) \,.
\label{eq:frames}
\end{equation}

The three-form potential is, imposing spherical symmetry on the $\mathbb{R}^4$s :

\begin{equation}
C^{(3)} ~=~   - e^0 \wedge e^1 \wedge e^2 ~+~   \Big((\partial_z w )^{-1}\, (\partial_{u} w)u^3 \,d\Omega_{u,\,3}~+~ (\partial_{v} w)  \, v^3d\Omega_{v,\,3} \Big)  \,,
 \label{eq:C3 def}
\end{equation}

This is a supersymmetric solution preserving eight supersymmetries provided $e^{A_0}$ and $w$ are derived from a pre-potential that we call $G$ and which is constrained to satisfy a Monge-Ampère like equation:

\begin{equation}
    \mathcal{L}_v G~=~\mathcal{L}_u G ~\partial_{zz} G~-~(\partial_z\partial_u G)^2~\,,
\end{equation}
where $\mathcal{L}_{u/v}$ are the spherically symmetric Laplacian on the two $\mathbb{R}^4$s. The metric functions are derived from the pre-potential through:
\begin{equation}
w ~=~ \partial_z G  \,, \qquad  e^{-3  A_0} \, (-\partial_z w )^{\frac{1}{2}} ~=~ {\cal L}_{v}  G \,.
 \label{eq:defAw}
\end{equation}


\subsection{D3-D5-NS5 intersection in type IIB}


An analogous D3-D5-NS5 intersection can be obtained from \eqref{eq:11metric} by first relaxing the assumption of spherical symmetry and imposing translational isometries along one of the $u_i$ and one of the $v_i$ directions. Reducing along the former and T-dualizing along the latter yields the following brane configuration \cite{Bena:2025hxt}:\\

\begin{center}
\begin{tabular}{c|cccc}
 & $x^{0,1,2}$ & $z$ & $\mathbb{R}_u^3$ & $\mathbb{R}_v^3$\\\hline
 D3 & $\times$ & $\times$ & &\\
 D5 & $\times$ & & $\times$ &\\
 NS5 & $\times$ & & & $\times$
\end{tabular}
\end{center}
With rotational symmetry in each $\mathbb{R}^3$s, the string-frame metric and fields take the form \cite{Lunin:2007mj,Bena:2025hxt}
\begin{align}
 \dd s_{10,s}^2={}&e^{3A_0/2}(-\partial_z w)^{-1/4}
 \left(\eta_{ab}\dd x^a\dd x^b+(-\partial_z w)\left(\dd z+(\partial_z w)^{-1}\partial_u w\,\dd u\right)^2\right)
\nonumber\\
 &+e^{-3A_0/2}(-\partial_z w)^{-1/4}
 \left((-\partial_z w)^{-1/2}\dd\vec u^{\,2}+(-\partial_z w)^{1/2}\dd\vec v^{\,2}\right),\nonumber\\
 e^{2\phi}={}&(-\partial_z w)^{-1},\qquad
 C_2=-v^2\partial_vw\,\vol(S_2^2),\qquad
 B_2=u^2\frac{\partial_uw}{\partial_zw}\,\vol(S_1^2),\nonumber\\
 C_4^{\rm el}={}&-e^{3A_0}(-\partial_z w)^{1/2}\dd x^0\wedge\dd x^1\wedge\dd x^2\wedge\left(\dd z+(\partial_z w)^{-1}\partial_u w\,\dd u\right).
 \label{eq:IIBbrane}
\end{align}
We wrote only the electric component of $C_4$; the magnetic component is determined locally by self-duality of the five-form field strength.

As in the M-theory construction, all fields are determined by a pre-potential $G$, which satisfies a Monge-Ampère-like equation of the same form:
\begin{equation}
 w=\partial_zG,\qquad
 \Lv G=(-\partial_z w)^{1/2}e^{-3A_0},\qquad
 \Lv G=(\partial_z^2G)\Lu G-(\partial_u\partial_zG)^2,
 \label{eq:def func IIB}
\end{equation}
where $\mathcal{L}_{u/v}$ denote the radial Laplacians on the
corresponding $\mathbb{R}^3$. The pre-potential giving rise to
$\AdS_5\times S^5$ is:
\begin{equation}
 G^{(0)}=-\frac{z^2}{2}-\frac{8}{uv}\arctan\frac{v}{u}\,,
 \qquad w=-z\,,\qquad e^{-3A_0}=H=\frac{16}{r^4}\,.
 \label{eq:IIBvac}
\end{equation}
The superscript $^{(0)}$ denotes the undeformed background, about
which we will expand the Janus solutions. Our normalization sets
the radius to $L^4=16$. Away from the sources, the pre-potential
satisfies $\Lv G^{(0)}=-\Lu G^{(0)}=H$.

The solution preserves the supersymmetry projectors of the
corresponding flat-brane intersection. In particular, a probe
D3-brane extended along $(x^a,z)$ at fixed $(u\,, v)$ feels
no force. To see this, consider its Dirac-Born-Infeld and
Wess-Zumino actions:
\begin{equation}
    S_{DBI}=-T_{D3}\int d^4\sigma e^{-\phi} \sqrt{-{\rm det}\left( g_{\alpha \beta}+\mathcal{F}_{\alpha\beta}\right)}-T_{D3} \sum_p \int d^4\sigma \,e^{\mathcal{F}} \wedge \tilde{C}_{p}\,,
\end{equation}
where $g_{\alpha \beta}$ is the induced metric on the D3-brane, $\mathcal{F}$ is the gauge invariant combination of the two form field strength on the worldvolume and the pull back of the $B$-field, and $\tilde{C}_{p}$ is the pull back of the $p$-form gauge field.  At fixed $(u,\,v)$, the DBI and electric Wess-Zumino contributions cancel, so the probe feels no force: the brane is floating. As we will see, it does not imply that the canonically normalized scalars describing transverse fluctuations on $\mathbb{R}^6=\mathbb{R}^3_u\times \mathbb{R}^3_v$ are massless.

Similarly, an NS5-brane extended along $(x^a,\mathbb{R}^3_v)$ at fixed $(z,u)$ feels no force. A floating D5-brane instead extends along $(x^a,\mathbb{R}^3_u)$ at fixed $(v,w)$. To evaluate its probe action, it is convenient to use $w$ as a coordinate and regard $z$ as a function of $(u,v,w)$.


\section{Janus interface in type IIB}
\label{sec:Janus interface in type IIB}

The supersymmetric type IIB Janus solution \cite{DHoker:2007hhe,DHoker:2007mci,DHoker:2007zhm} is dual to an interface across which the Yang-Mills coupling jumps. The interface action contains a localized fermion bilinear and its supersymmetric scalar completion \cite{DHoker:2006qeo,Gaiotto:2008sd}. We interpret this bilinear as a localized mass deformation. To identify its gravitational counterpart, we express the exact Janus solution in brane coordinates and show that, to first order in the Janus parameter, the two-form potential is the bulk response to a delta-function source giving equal masses to the four Weyl fermions. Smearing this perturbation reproduces the corresponding Polchinski-Strassler flux. The accompanying dilaton gradient is essential to supersymmetry: the smeared configuration retains a varying Yang-Mills coupling, whereas the homogeneous deformation giving equal masses to all four fermions at fixed coupling is not supersymmetric.


\subsection{Field theory: the Janus interface as a localized mass}


Supersymmetric planar interfaces in $\mathcal{N}=4$ super Yang-Mills have been classified under the requirement that they preserve the three-dimensional Lorentz and conformal symmetries of the interface \cite{DHoker:2006qeo}. We begin by fixing conventions for $\mathcal{N}=4$ super Yang-Mills. All fields transform in the adjoint of $su(N)$, and we suppress their gauge indices. The bosonic fields consist of a gauge field $A_\mu$, which is a singlet under the $SO(6)$ R-symmetry, and six real scalars $\Phi_I$ transforming in its $\bf 6$. We organize the scalars into two triplets, $U^{I},V^{I}$ with $I=1,2,3$.The fermionic degrees of freedom are four Weyl fermions $\lambda^a$ in the $\bf 4$.

We follow the conventions of D'Hoker, Estes and Gutperle \cite{DHoker:2006qeo}. We introduce the six matrices $\rho^I$ following there notations, they couple the $SU(4)$ indices of two fermions to the scalar vector representation, while the charge-conjugation matrix $\mathcal{C}$ contracts spacetime spinor indices:
\begin{equation}
 (\rho^I)^T=-\rho^I\,,\qquad
 (\rho^I)^*\rho^J+(\rho^J)^*\rho^I=-2\delta^{IJ}\mathbf1_4\,,
 \qquad
 (\Gamma^\mu)^T=-\mathcal C\Gamma^\mu\mathcal C^{-1}\,.
 \label{eq:SYMspinorconventions}
\end{equation}
Thus $\mathcal C$ has no $SO(6)$ index. The trace acts only on $su(N)$ indices. With covariant derivative $D_\mu=\partial_\mu+i[A_\mu,\,\cdot\,]$ the Lagrangian is:
\begin{align}
    S\!=\!-\!\int \!\!d^4x& T_{D3}\text{Tr}\left(\frac{1}{2}F_{\mu \nu}F^{\mu \nu}+ D_{\mu}U^ID^{\mu}U^I+D_{\mu}V^ID^{\mu}V^I+\frac{1}{2}[V^I,V^J][V^I,V^J]+\frac{1}{2}[U^I,U^J][U^I,U^J]\right)\nonumber\\
    &+\text{Tr}\left([V^I,U^J][V^I,U^J]-i\bar\lambda\Gamma^{\mu}D_{\mu}\lambda+i D_{\mu} \bar{\lambda}\Gamma^{\mu}\lambda+\lambda^T \mathcal{C}\rho^I[U^I,\lambda]+\lambda^{\dagger}\mathcal{C}(\rho^{I})^{*}[U^I,\lambda^{*}]\right)\nonumber\\
    &+\text{Tr}\left(\lambda^T \mathcal{C}\rho^I[V^I,\lambda]+\lambda^{\dagger}\mathcal{C}(\rho^{I})^{*}[V^I,\lambda^{*}]\right)\,.
\end{align}

where $T_{D3}=\frac{1}{2 g^2}$. An interface with varying Yang-Mills coupling $g(\tilde{z})$, where $\tilde{z}$ is one of the worldvolume directions, is consistent with conformal invariance only if it is a step function (zero width). Nonetheless it has to be thought as the limit of a smoothly varying $g_{YM}$. The maximally supersymmetric interface which preserves conformal and Poincaré invariance in the $2+1$ worldvolume directions transverse to the interface is:
\begin{equation}
    S_{\text{Int}}=\int d^4x \frac{\partial_{\tilde{z}} g}{g^3}\text{Tr}\left(\frac{i}{2} \lambda^T\mathcal{C}\lambda+\frac{i}{2} \lambda^{\dagger}\mathcal{C}\lambda^{*}-2 i\epsilon_{ijk} U^{I}[U^{J},U^{K}]+\partial_z\left(U^IU_I\right)-\frac{2 \partial_z g}{g} \left(U^IU_I\right)\right)\,.
\end{equation}
The distributional limit of each term need not exist as long as the overall limit does exist. The term which looks like a boson mass may be removed by a redefinition of the $U^I$ scalars.  With $\tilde{U}^I=\frac{U^I}{g^2}$, the Lagrangian of the interface can be written :
\begin{equation}
 S_{\text{Int}}=\int d^4x
 \frac{\partial_{\tilde{z}} g}{g^3}\tr\left(
 \frac i2\lambda^TC\lambda+\frac i2\lambda^\dagger C\lambda^*
 -2ig^6\epsilon_{IJK}U^I[U^J,U^K]
 \right)\,.
 \label{eq:IIBinterface}
\end{equation}
After canonical normalization the fermion mass is proportional to $\partial_{\tilde{z}}\log g$, and the quadratic bosonic interaction can be removed from the field redefinition $\hat{U}^I=\frac{\hat{U}^I}{g^2}$ (the kinetic term of the scalar fields is not canonically in these variables). The same coefficient multiplies a cubic scalar interaction $\partial_{\tilde{z}}\log g$, see \cite{DHoker:2006qeo} for more details. For a step coupling,
\begin{equation}
 \partial_{\tilde{z}}\log g=\log\left(\frac{g_+}{g_-}\right)\,\delta(\tilde{z}).
\end{equation}
The interface therefore contains a localized equal-mass fermion bilinear together with its supersymmetric completion. In this section we show how the flux in the exact geometry realizes precisely this localized mass deformation, we study this using techniques introduced by Polchinski and Strassler.

The deformation preserves eight Poincar\'e supercharges for smooth $g(\tilde{z})$. The conformal interface also preserves eight conformal supercharges and therefore has $OSp(4|4)$ symmetry. It predicts a localized three-form in the equal-mass harmonic, a supersymmetric no-force D3 embedding, a relative scalar rescaling across the interface, and a varying dilaton. The last ingredient is indispensable when comparing with a homogeneous mass at fixed coupling.


\subsection{Polchinski-Strassler in brief}
\label{sec:PolchinskiBrief}

Polchinski and Strassler studied mass deformations of $\mathcal{N}=4$ super Yang-Mills perturbatively in supergravity \cite{Polchinski:2000uf}. We briefly review their construction, which will guide our analysis of localized mass deformations. An analogous treatment for the M2-brane field theory was developped in \cite{Bena:2000zb}. We consider perturbations around $AdS_5\times S^5$ and introduce the complex three-form field strength:
\begin{equation}
      G_{3} = F_{3} - \tau H_{3},.
\end{equation}
This field strength vanishes in the undeformed background and is first order in the mass parameter. Here $\tau = C_0 + i e^{-\Phi_0}$ is the background axio-dilaton. Its perturbation contributes to $G_{3}$ only at second order, since $H_{3}$ is itself first order. In our conventions, the background value is $\tau=i$. The linearized equation of motion and Bianchi identity are
\begin{equation}
{\rm d} (* G_{3}) + i G_{3} \wedge F_{5,\,\AdS}=0\quad {\rm d} (G_{3}) = 0\, .
\end{equation}
We restrict attention to three-form perturbations with legs entirely along the six directions transverse to the D3-branes, parametrized by $(u_1,u_2,u_3)$ and $(v_1,v_2,v_3)$. For such perturbations, the ten-dimensional Hodge dual takes the form
\begin{equation}
*G_{ 3} =  Z^{-1} {*_6} G_{ 3} \wedge
dx^0 \wedge dx^1 \wedge dx^2 \wedge  dx^3 \,.
\end{equation}
Here $*_6$ is the Hodge star for the flat transverse metric, and $Z=\frac{1}{(u^2+v^2)^2}$ is the D3-brane warp factor in units where the AdS radius is one (only for this review part). The three-form equation of motion then becomes:
\begin{equation}
{\rm d} \left( \frac{1}{Z} ( *_6 G_{3} - i  G_{3})\right)= 0\, .
\end{equation}
It is convenient to introduce complex coordinates:
\begin{equation}
z^i = \frac{v_i + i ~u_i}{\sqrt 2}.
\end{equation}
The fermion bilinears transform under the $SO(6)$ R-symmetry of $\mathcal{N}=4$ super Yang-Mills which are  the $\bf 10$ and $\bar{\bf 10}$. These representations can be realized as imaginary self-dual and imaginary anti-self-dual antisymmetric three-tensors in the transverse six-dimensional space. To write them explicitly, we use $x^a=u_a$ for $a\in [1,3]$ and $x^{a+3}=v_a$ for $a\in [1,3]$:
\begin{equation}
*_6 T_{ijk} \equiv \frac{1}{3!} \epsilon_{ijk}{}^{lmn} T_{lmn} = \pm ~i~ T_{ijk} \,,\label{dual}
\end{equation}
where the $+$ refers to  ${\bf 10}$ while the $-$ stands for $\bar{\bf 10}$. Following the notations of Polchinski and Strassler we define:
\begin{equation}
V_{ijk} = \frac{x^q}{u^2+v^2} ( x^i T_{qjk} + x^j T_{iqk} + x^kT_{ijq})
\end{equation}
and the quantities :
\begin{eqnarray}
T_{ 3} &~=~ \frac{1}{3!}  T_{ijk} dx^i\wedge dx^j\wedge dx^k \,,\nonumber\\
V_{ 3} &~=~ \frac{1}{3!}  V_{ijk}dx^i\wedge dx^j\wedge dx^k \,,
\nonumber\\
S_{2}&~=~ \frac{1}{2} T_{ijk} x^i dx^j\wedge dx^k\,.
\end{eqnarray}
We highlight the following useful identities:
\begin{align}
dS_{2} ~=~& 3T_{ 3}\ ,\quad {\rm d}(\ln r)\wedge S_{ 2} ~=~ V_{ 3}\\
{\rm d}(r^p S_{2})& ~=~ r^p(3 T_{ 3} + p V_{ 3})\nonumber\\
{\rm d}T_{ 3}~=~&0\,,\quad {\rm d}V_{ 3} = -3 {\rm d}(\ln r) \wedge T_{3}\,,
\end{align}
and
\begin{equation}
{*_6} T_{ 3} = \pm i T_{ 3}\ ,\quad
{*_6} V_{3} = \pm i (T_{ 3}-V_{ 3})\,. \label{uvdual}
\end{equation}
Given the transformation of the fermions under a phase rotation $z_i\rightarrow e^{i\theta_i}z_i$, we obtain that the mass deformation of the fermions leads to :
\begin{equation}
T_{ 3} = m_{1} dz^1 \wedge d\bar z^2 \wedge d\bar z^3 +
m_{2} d\bar z^1 \wedge dz^2 \wedge d\bar z^3 +
m_{3} d\bar z^1 \wedge d\bar z^2 \wedge dz^3 +
m_{4} dz^1 \wedge dz^2 \wedge dz^3 \,.
\end{equation}


\subsection{The super-conformal Janus solution as a near brane limit}


Taking a near horizon limit of the solution of the section \ref{sec:Brane intersections} can enlarge the symmetry group of the solution as is familiar in AdS/CFT. In particular, large families of solutions with isometry group $SO(2,3)\times SO(3)\times SO(3)$ arise from brane intersections\cite{Bena:2023rzm,Bena:2024dre,Bena:2025hxt}. The super-conformal Janus interface has exactly these isometries.

In this section we explain the construction of the Janus solutions found in \cite{DHoker:2007hhe,DHoker:2007zhm}, focusing on the conformal interface. The metric of the Janus solution in type IIB supergravity is a fibration of $AdS_4\times S^2\times S^2$ over a Riemann surface with boundaries (taken to be the infinite strip) parametrized by complex coordinates $(\zeta,\bar{\zeta})$. The functions are parametrized by two positive harmonic functions on the Riemann surface $h_1$ and $h_2$.
\begin{equation}\label{tenmet2}
    d s^2 = f_4^2 d s_{{\rm AdS}_4}^2 + f_1^2 ds_{{\rm S}_1^2}^2 + f_2^2 ds_{{\rm S}_2^{2}}^2 + 4 \rho^2 d\zeta d\bar \zeta\ ,
\end{equation}

where the metric coefficients are defined by:
\begin{align}
    f_4^8 &= 16\, {N_1 N_2 \over W^2}\ ,
\qquad \qquad
\rho^8 = \frac{N_1 N_2 W^2 }{ h_1^4 h_2^4}\ ,  \nonumber \\
f_1^8 &=  16\,  h_1^8 \frac{N_2 W^2 }{ N_1^3}\ ,
\qquad  \qquad
f_2^8 = 16\,  h_2^8 \frac{N_1 W^2}{N_2^3}\,.
\end{align}
The functions $W,\,N_1, \,N_2$ are given in terms of the harmonic functions $h_1$ and $h_2$ by:
\begin{align}
    W &= \partial  h_1 \bar\partial  h_2 + \bar\partial h_1 \partial  h_2  = \partial \bar\partial  (h_1h_2)\ ,  \nonumber \\
N_1 &= 2 h_1 h_2 |\partial h_1|^2 - h_1^2 W \ , \nonumber \\
N_2 &= 2 h_1 h_2 |\partial h_2|^2 - h_2^2 W   \ ,
\label{eq:harmonic function}
\end{align}
The Riemann surface is taken to be the infinite strip with real $(x,y)$ coordinates defined as: $\zeta=x+i y$,with $x\in ]-\infty,+\infty[ $ and $y\in [0,\frac{\pi}{2}]$, the boundary of the Riemann surface are at $y=0,\,\frac{\pi}{2}$. Janus interface is a one parameter deformation of $AdS_5\times S^5$ with harmonic functions given by:
\begin{align}
\label{eq:harmonicIIB}
    h_1(x,y)= 2 \cos(y)\cosh(x+\delta)\,,\quad &h_2(x,y)= 2 \sin(y) \cosh(x-\delta)\\
    h_{1D}(x,y)= 2 \sin(y) \sinh(x+\delta) \,,\quad &h_{2D}(x,y)=-2 \cos(y)\sinh(x-\delta)\,.
\end{align}
$AdS_5\times S^5$ is simply obtained by taking the $\delta=0$ limit:
\begin{align}
    h_1(x,y)= 2 \cos(y)\cosh(x)\,,\quad &h_2(x,y)= 2 \sin(y) \cosh(x)\nonumber \\
    h_{1D}(x,y)= 2 \sin(y) \sinh(x) \,,\quad &h_{2D}(x,y)=-2 \cos(y)\sinh(x)\,.
    \label{eq:Janus solution}
\end{align}
Since we want to give an interpretation of the solution as the zooming in of branes in flat space, it is natural to write $AdS_4$ in Poincaré coordinates:
\begin{equation}
    ds_{{\rm AdS}_4}^2=\frac{\d\mu^2}{\mu^2}+\mu^2\left(-dt^2+dx_1^2+dx_2^2\right)\,.
\end{equation}

These solutions are describing the near brane limit with enhanced symmetry of $\frac{1}{4}$-BPS intersection of branes. The map between the brane \eqref{eq:IIBbrane} is expressed in terms of the harmonic functions in \eqref{eq:harmonicIIB} through the identification:
\begin{equation}
    u~=~ 2 \,\mu \, h_1\,,\qquad v~=~2\, \mu\, h_2\,,\qquad z~=~\frac{h_2^D}{\mu \,h_1}  \,, \qquad w~=~\frac{h_1^D}{\mu \,h_2}\,.
\label{eq:UVmap}
\end{equation}
For Janus the precise matching is obtained by:
\begin{align}
     u= 4 \mu \cos(y)\cosh(x+\delta)\,,\quad &z=-\frac{\sinh(x-\delta)}{\mu\cosh(x+\delta)}\\
     v= 4 \mu \sin(y) \cosh(x-\delta)\,,\quad & w=\frac{\sinh(x+\delta)}{\mu\cosh(x-\delta)}\,.
     \label{eq:def uvz}
\end{align}
To obtain $w$ as a function of $(u,v,z)$, eliminate $\mu$ and $y$ from \eqref{eq:UVmap}. The last strip coordinate $x=x(u,v,z,\delta)$ is determined through:
\begin{equation}
 z^2\left(u^2\cosh^2(x-\delta)+v^2\cosh^2(x+\delta)\right)
 =4\sinh^2(2x-2\delta).
 \label{eq:IIBimplicit}
\end{equation}
This is a quartic equation for $e^{2x}$. We define $r^2=u^2+v^2$. Choose the root with
$\operatorname{sgn}\sinh(x-\delta)=-\operatorname{sgn}z$, continued from
$\sinh x=-\frac{rz}{4}$ at $\delta=0$. Then:
\begin{equation}
 w(u,v,z)=-z\,\frac{\sinh(2x+2\delta)}{\sinh(2x-2\delta)}\,.
 \label{eq:IIBexactw}
\end{equation}
In practice it is impossible to obtain the full solution $G_0(u,v,z)$ from which the Janus solution is derived in the brane in flat space coordinates. We can choose $\delta$ to be a small parameter and work in perturbation theory. It will be enough to capture most of the physics of the solution. We therefore expand $G=G^{(0)}+\delta G^{(1)}+\delta^2G^{(2)}+o(\delta^2)$. The first two corrections follow :
\begin{align}
 G^{(1)}&=\frac{2z}{r}\sqrt{16+r^2z^2},\nonumber\\
 G^{(2)}&=-\frac{4[64+4(5u^2+3v^2)z^2+r^4z^4]}{r^2(16+r^2z^2)}
 -\frac{16}{uv}\arctan\frac vu.
 \label{eq:IIBseries}
\end{align}
The first-order pre-potential is odd in $z$ and depends on the transverse coordinates only through $r$. The angular dependence appears at second order. The final term in $G_0^{(2)}$ adjusts the $\AdS_5$ radius.


\subsection{The localized mass}


The pre-potential determines the first-order perturbation of the dilaton:
\begin{equation}
 e^{2\phi}=1+\delta\frac{4rz(24+r^2z^2)}{(16+r^2z^2)^{3/2}}+O(\delta^2).
\end{equation}
The transition has a width of order $r^{-1}$ and becomes sharp as the transverse radius increases. At fixed nonzero $z$, the limit $r\rightarrow \infty$ gives
\begin{equation}
 e^{2\phi}=1+4 ~\delta ~\text{sign} (z)+O(\delta^2).
\end{equation}
This is the expected dilaton profile for an interface across which the Yang-Mills coupling $g$ jumps. At first order in $\delta$, the two-form is:
\begin{equation}
 C_2-iB_2=\frac{512\delta}{r^3(16+r^2z^2)^{3/2}}S_2+O(\delta^2)\,.
 \label{eq:IIBflux}
\end{equation}
To identify the source profile, we work to zeroth order in the coordinate map, for which the Poincar\'e radial coordinate is $\nu=r/4$ and $\tilde{z}=z$. For a source localized on the interface and constant along it, the bulk response is obtained by integrating the bulk-to-boundary propagator of a dimension-three operator over the interface:
\begin{equation}
 \int_{\R^3}\dd^3x_\parallel
 \left(\frac{\nu^{-1}}{\nu^{-2}+\tilde{z}^2+|x_\parallel|^2}\right)^3
 =\frac{\pi^2}{4(1+\nu^2\tilde{z}^2)^{3/2}}.
 \label{eq:IIBkernel}
\end{equation}
After factoring out the sphere harmonic, the remaining radial dependence agrees with~\eqref{eq:IIBflux}, identifying the perturbation as the response to a mass source localized on the interface. A homogeneous source would instead require integrating the propagator over $\tilde{z}$ as well.

Smearing the localized source in the transverse worldvolume direction makes the comparison with Polchinski-Strassler explicit:
\begin{align}
 \int_{-\infty}^{\infty}\dd z\,(C_2-iB_2)&=\frac{64\delta}{r^4}S_2,\nonumber\\
 G_3^{\rm smear}&=\frac{192\delta}{r^4}\left(T_3-\frac43V_3\right).
 \label{eq:IIBsmear}
\end{align}
With $T_{3}=\frac{1}{3!}\varepsilon_{ijk}\left(dv^i\wedge dv^j\wedge dv^k-i~du^i\wedge du^j\wedge du^k\right)\,,$ we identify that the deformation corresponds to one where all the masses of the fermions are equal. In the $\mathcal{N}=1$ formulation of $\mathcal{N}=4$ super Yang-Mills it means that we gave the same mass to the fermions in the hyper-multiplet as the one in the vector multiplet. Giving a mass to the gluino (the fermions in the vector multiplet), leads to a $\mathcal{N}=0$ theory. 

This appears to create a puzzle: at first order, smearing the supersymmetric interface perturbation preserves the linearized Monge-Ampère-like equation, yet reproduces the flux associated with a nonsupersymmetric homogeneous mass deformation. The resolution is that the naïve smearing of the dilaton

\begin{equation}
    \int_{-\infty}^{\infty}\dd z \exp(2\phi)=1\,,
\end{equation}
is inconsistent. One must solve for the Monge-Ampère equation to first order in $\delta$ which leads to $G_3^{\rm smear}$, and then compute the dilaton. The pre-potential can be written:
\begin{equation}
 G_{\text{ smeared}}=G^{(0)}+\frac{4\delta}{3}\left(z^3+\frac{24z}{r^2}\right)+O(\delta^2)\,.
\end{equation}
It yields:
\begin{equation}
 C_2-iB_2=\frac{64\delta}{r^4}S_2,
 \qquad e^{4\phi}=1+16\delta z+O(\delta^2)\,.
\end{equation}
Of course the approximation $\delta \ll1$ fails at $z\sim \frac{1}{\delta}$. We keep this section brief since this duality is already well known, we only shed  new light and in particular new physical intuition on why the duality is actually true besides all the checks obtained for example in \cite{Karch:2026ymg,Komatsu:2020sup,He:2025due,He:2026qnv}. It also serves as a guiding principle to study the newest results of section \ref{sec:conformalMassdef}.


\subsection{The six massless scalars and the gluing condition}
\label{sec:The six massless scalars}

We have shown that a probe D3 brane whose worldvolume is along $(x^0,x^1,x^2,z)$ is floating (and hence that the solution have the same killing spinor as that of a flat D3-brane along these directions). Nonetheless (and as we will mostly see in the mass dependent interface), it is still non trivial to check that the transverse scalar fields remain massless. In this subsection we therefore expand the probe action around a fixed $(u,v)=(u_0,v_0)$, and explicitly show that the scalar fields remain massless in the dual field theory, and that the gluing conditions along the interface are reproduced.

Let us consider a D3 along $(x^0,x^1,x^2,z)$ and expand about a fixed position in both transverse three-planes,
\begin{equation}
  u= u_0+~\epsilon~ u(x^a,z),\qquad
 v=v_0+~\epsilon~  v(x^a,z).
 \label{eq:D3fluctuations}
\end{equation}
And we expand the Born-Infeld plus Wess-Zumino action to quadratic order in $\epsilon$ . We keep the worldvolume gauge field zero. The pulled-back $B_2$ starts at second order in these displacements and contributes to the determinant only at fourth order. The other Wess-Zumino terms likewise do not enter the quadratic scalar action.

It is useful to keep the fibration in~\eqref{eq:IIBbrane} until after expanding the determinant. If the explicit transverse metric terms are omitted, its pullback gives exactly:
\begin{equation}
 e^{-\phi}\sqrt{-\det g_{\rm ind}}
 =e^{3A_0}(-\partial_z w)^{1/2}
 \left(1+(\partial_z w)^{-1}\partial_u w\,\partial_z u\right).
 \label{eq:D3electriccancel}
\end{equation}
The electric four-form has precisely the same factor, with the opposite contribution to the action. Thus all terms without transverse derivatives, including their Taylor expansion around $(u_0, v_0)$, cancel. The term linear in $\partial_z u$ cancels as well: the D3-brane at $\epsilon=0$ is floating. Therefore the expansion leads to:
\begin{align}
 S_{\rm D3}^{(2)}=-\frac{T_3\epsilon^2}{2}\int\dd^3x\,\dd z\,
 \Big[&\eta^{ab}\partial_a u\cdot\partial_b u
 +\frac{(\partial_z u)^2}{-\partial_z w}\nonumber\\
 &+(-\partial_z w)\eta^{ab}\partial_a v\cdot\partial_b v
 +(\partial_z v)^2\Big].
 \label{eq:D3quadratic}
\end{align}
Here $\partial_z w$ is evaluated to first order in the Janus parameter $\delta$. To put the worldvolume kinetic terms in Lorentz-invariant form, we introduce the coordinate $\tilde{z}$ through $d\tilde{z}=(-\partial_zw)^{\frac{1}{2}}dz$, evaluated on the fixed probe embedding. We identify $\tilde{z}$ with the transverse coordinate of the field theory. This choice treats the two brane-coordinate descriptions symmetrically: along the probe, $dw=(\partial_zw)dz$, so $d\tilde{z}$ is the geometric mean of $dz$ and $dw$. At zeroth order in $\delta$, $\tilde{z}$ coincides with $z$, up to an additive constant. We also introduce the canonically normalized scalar fields:
\begin{equation}
    \tilde{U}=(-\partial_z w)^{-\frac{1}{4}}u\,,\quad \tilde{V}=(-\partial_z w)^{\frac{1}{4}}v\,.
\end{equation}
It then follows that the effective action is given by:
\begin{align}
 S_{\rm D3}^{(2)}=-\frac{T_3\epsilon^2}{2}\int\dd^3xd\tilde{z}\,
 \Big[&\eta^{ij}\partial_i \tilde{U}\cdot\partial_j \tilde{U}
 -\delta\frac{1920 r_0^3\tilde{z}}{(16+r_0^2\tilde{z}^2)^{\frac{7}{2}}}\tilde{U}^2\nonumber\\
 &+\eta^{ij}\partial_i \tilde{V}\cdot\partial_j \tilde{V}
 +\delta\frac{1920 r_0^3\tilde{z}}{(16+r_0^2\tilde{z}^2)^{\frac{7}{2}}}\tilde{V}^2\,,
 \label{eq:D3quadratic}
\end{align}
with $\eta^{ij}$ the four dimensional Lorentz invariant metric and $r_0^2=u_0^2+v_0^2$. The boundary of $\AdS_5\times S^5$ is at $r_0\rightarrow \infty$, the boson mass can be identified as the coefficient in front of the quadratic part of $\tilde{U},\,\tilde{V}$. For Janus interface, and as predicted by the field theory, the mass of the bosons vanishes. But, as we will show now, it vanishes into a non trivial distributional value.

Let us take the interface Lagrangian \eqref{eq:IIBinterface} following \cite{DHoker:2006qeo}, after integration by parts, the quadratic part of the action is given by:
\begin{equation}
    S=-\int d^4x\frac{1}{2g^2} \left(D_{\mu}U^ID_{\mu}U^I+D_{\mu}V^ID_{\mu}V^I+\left(4\frac{(\partial_{\tilde{z}}g)^2}{g^2}+2g^2\partial_{\tilde{z}}\left(\frac{\partial_{\tilde{z}}g}{g^3}\right)\right)U^IU^I\right)\,.
\end{equation}
After rescaling of the six bosons through $\tilde{X}^I=\frac{X^I}{g}$ \footnote{This scaling is different from the one of \cite{DHoker:2006qeo} which was : $\hat{U}^I=\frac{\hat{U}^I}{g^2}$, for which the bosonic mass deformation vanishes, at the cost of a non-canonically normalized boson.}(to have a canonically normalized action), and integration by parts, one obtains:
\begin{equation}
    S=-\int d^4x\frac{1}{2} \left(D_{\mu}\tilde{U}^ID_{\mu}\tilde{U}^I+D_{\mu}\tilde{V}^ID_{\mu}\tilde{V}^I+\left(\frac{\partial_{\tilde{z}\tilde{z}}g}{g}\right)\tilde{U}^I\tilde{U}^I+\left(2\left(\frac{\partial_{\tilde{z}}g}{g}\right)^2-\frac{\partial_{\tilde{z}\tilde{z}}g}{g}\right)\tilde{V}^I\tilde{V}^I\right)\,.
\end{equation}
And defining $m(\tilde{z})=-\frac{\partial_{\tilde{z}}g}{g}$ we can identify \footnote{Similar expressions can be found in \cite{Arav:2020obl}.}: 
\begin{equation}
    M_U^2=m(\tilde{z})^2-m'(\tilde{z})\,,\quad M_V^2=m(\tilde{z})^2+m'(\tilde{z})\,.
\end{equation}
These are the masses to compare with what we obtained in the probe computation of \eqref{eq:D3quadratic}, and we obtain:
\begin{equation}
    m(\tilde{z})=-\frac{384 \delta r_0}{\left(16+r_0^2\tilde{z}^2\right)^{\frac{5}{2}}}\,.
\end{equation}
This mass integrate non trivially for any $r_0$ to 
\begin{equation}
    \int_{-\infty}^{+\infty}m(\tilde{z})d\tilde{z}=-2\delta\,,
\end{equation}
and therefore $m(\tilde{z})$ converges distributionally to $m(\tilde{z})=-2\delta~ \delta(\tilde{z})$
Therefore in the large $r_0$ limit, we obtain a gluing condition:
\begin{equation}
\label{eq:IIBgluingcond}
    \tilde{U}(0^{+})~=~(1+2\delta)~\tilde{U}(0^{-})\,,\quad  \tilde{V}(0^{+})~=~(1-2\delta)~\tilde{V}(0^{-})\,.
\end{equation}
This is what is expected from the field theory, indeed, assume all the $U^I,\,V^I$ to commute, the supersymmetric equations are:
\begin{equation}
    \partial_{\tilde{z}}\tilde{U}=-m(\tilde{z}) \tilde{U}\,,\quad\partial_{\tilde{z}}\tilde{V}=+m(\tilde{z}) \tilde{V}\,.
\end{equation}
For a delta function source of strength $2\delta$, it leads to:
\begin{equation}
    \tilde{U}(0^{+})~=~e^{2\delta}~\tilde{U}(0^{-})\,,\quad  \tilde{V}(0^{+})~=~e^{-2\delta}~\tilde{V}(0^{-})\,.
\end{equation}
To first order we recover \eqref{eq:IIBgluingcond}.

\section{Janus interface in M-theory}
\label{sec:Janus interface in Type Mtheory}

This section is devoted to reproducing the analysis of the previous section for the Janus conformal interface in M-theory and introduce notations for the last section of this paper.


\subsection{ABJM and its supersymmetric mass deformations}


The starting point is the conformal field theory on $N$ coincident M2-branes. It is described by ABJM theory which is a $U(N)_k\times U(N)_{-k}$ Chern-Simons theory coupled to four complex bifundamental scalars $Y^A$ and their fermionic partners $\psi_A$ \cite{Aharony:2008ug,Aharony:2008gk}. The scalar kinetic term is $-\tr(D_\mu Y_A^\dagger D^\mu Y^A)$. The undeformed scalar interaction is sextic, as required by the dimension $\frac{1}{2}$ of $Y^A$. We consider only $k=1$, for which the theory describes $N$ coincident M2-branes in flat transverse space, with no orbifold quotient. It is useful first to describe the deformation in the usual $\mathcal{N}=6$ ABJM variables. We will later focus on mass deformations of this theory, and in particular, its space-time varying mass deformations. For this last purpose, we will follow the construction of Kim and Kwon and the study of its vacua by Kim, Kim, Kwon and Kim \cite{Kim:2018qle,Kim:2019kns}. We also use there conventions for the Lagrangian of ABJM and supersymmetry variations. We first introduce a constant supersymmetric mass, and then allow it to depend on one spatial coordinate. The second step requires an additional scalar interaction. This interaction is central to both the Janus interface and the conformal mass profile studied in the next section.

We will shortly introduce the operators corresponding to a spacetime varying mass, but before that, let us write the mass matrix and label the scalar fields as:

\begin{equation}
 Y^A=(V^1,V^2,U^1,U^2),\qquad
 M^A{}_B=\operatorname{diag}(1,1,-1,-1).
 \label{eq:FTquartets}
\end{equation}

The two complex $U$ fields describe the four real $u$ directions, and the two complex $V$ fields describe the four real $v$ directions. The fermions in the two sectors have opposite signed mass parameters. This choice fixes what we mean by the sign of $m$.

For a constant $m$, supersymmetry adds three terms to the ABJM Lagrangian: a fermion bilinear proportional to $m$, a quartic scalar interaction proportional to $m$, and a scalar mass proportional to $m^2$. To display their relation, use the operators:

\begin{align}
 \mathcal O_1={}&M^A{}_B\tr(Y_A^\dagger Y^B)
 =\tr(V^\dagger V-U^\dagger U),\nonumber\\
 \mathcal O_2={}&iM^A{}_B\tr(\psi^{\dagger B}\psi_A)
 -\frac{4\pi}{k}M_B{}^D\tr\left(
 Y_C^\dagger Y^C Y_D^\dagger Y^B
 -Y^C Y_C^\dagger Y^B Y_D^\dagger\right)\,.
 \label{eq:FTmassoperators}
\end{align}
The constant mass theory deformation of ABJM follows:
\begin{equation}
 \mathcal L_{\rm const}=\mathcal L_{\rm ABJM}
 -m~\mathcal O_2-m^2~\tr\left(Y_A^{\dagger}Y^{A}\right)\,.
\end{equation}
It preserves the twelve real supercharges visible in the $\mathcal N=6$ formulation. All eight real scalars have mass squared $m^2$, although the signed fermion masses differ between the two families of $\mathbb{R}^4$-scalars. The quartic interaction also permits non-commuting scalar configurations, which will be important for the vacuum structure in the last section of this paper.

Let us split the worldvolume coordinates between $\tilde{z}$ the direction normal to the interface we shall introduce shortly, and $(x^0,x^1)$ parallel to the interface. We introduce the mass dependent parameter $m=m(\tilde{z})$. If we simply make this replacement in $\mathcal L_{\rm const}$ , supersymmetry is broken since a derivative in the fermion kinetic term now acts on $m$. In the construction of \cite{Kim:2018qle,Kim:2019kns}, the mass-dependent fermion variation is:
\begin{equation}
 \delta_m\psi_A=mM_A{}^B\omega_{BC}Y^C\,,
\end{equation}

where $\omega_{AB}$ is the antisymmetric ABJM supersymmetry parameter, with its usual reality condition. Restrict it by:

\begin{equation}
 \Gamma^{\tilde{z}}\omega_{12}=-\omega_{12}\,,\qquad
 \Gamma^{\tilde{z}}\omega_{34}=-\omega_{34}\,,\qquad
 \Gamma^{\tilde{z}}\omega_{ai}=+\omega_{ai}
 \quad(a=1,2,\quad i=3,4)\,.
 \label{eq:FTprojection}
\end{equation}

These projections preserve half the manifest supercharges of the mass deformed theory. The remaining variation of the constant-mass Lagrangian, up to a total derivative, takes the simple form:

\begin{equation}
 \delta\mathcal L_{\rm const}
 =\partial_{\tilde{z}}(m(\tilde{z}))\,\delta\mathcal O_1\,.
\end{equation}

We now define $m(\tilde{z})'=\partial_{\tilde{z}}(m(\tilde{z}))$. It is canceled by adding $-m'\mathcal O_1$. The result is therefore
\begin{equation}
\mathcal L=\mathcal L_{\rm ABJM}-m(\tilde{z})\mathcal O_2
 -m(\tilde{z})^2\tr(Y_A^\dagger Y^A)-m'(\tilde{z})\mathcal O_1\,.
 \label{eq:Lm}
\end{equation}
This is the inhomogeneous mass deformation. It preserves six real Poincar\'e supercharges in the manifest ABJM formulation for any smooth profile $m(\tilde{z})$.

One direct physical consequence is that the boson mass differs in the $u-$plane and in the $v-$plane, restricting the lagrangian to the quadratic part of the scalar fields:
\begin{equation}
 V^{(2)}=(m^2-m')\tr(U^\dagger U)+(m^2+m')\tr(V^\dagger V)\,.
 \label{eq:FTquadraticgeneral}
\end{equation}

The derivative of the fermion mass splits the two families of scalar fields. This is why the scalar masses are not trivially obtained by squaring the local fermion mass alone. For commuting fields the corresponding normal-direction operators nicely factorize:

\begin{align}
 -\partial_{\tilde{z}}^2+m^2-m'&=(-\partial_{\tilde{z}}+m)(\partial_{\tilde{z}}+m)\,,\nonumber\\
 -\partial_{\tilde{z}}^2+m^2+m'&=(\partial_{\tilde{z}}+m)(-\partial_{\tilde{z}}+m)\,.
 \label{eq:susyqm}
\end{align}

Their supersymmetric first-order equations are:

\begin{equation}
 (\partial_{\tilde{z}}+m)U=0,\qquad (\partial_{\tilde{z}}-m)V=0\,.
 \label{eq:abelianBPS}
\end{equation}

These equations will also explain how a floating brane can correspond to a non-constant canonically normalized scalar.

Unlike the Yang-Mills coupling in the IIB example, the Chern-Simons level is quantized and cannot be varied smoothly across an ABJM interface. Here we instead localize the mass deformation. Take a smooth profile supported in a thin layer, with $\int d\tilde{z}\,m(\tilde{z})=m_0$, and integrate \eqref{eq:abelianBPS} across that interface. In the zero-width limit,
\begin{equation}
 V^{(+)}=e^{m_0}V^{(-)},\qquad U^{(+)}=e^{-m_0}U^{(-)}\,.
 \label{eq:jumpM}
\end{equation}

The superscripts denote limits on each side of the interface. It gives the commuting-sector supersymmetric gluing condition.

Away from the interface at $\tilde{z}=0$, clearly $m=m'=0$, so the bulk scalars and fermions have no mass deformation. The interface nevertheless changes their matching conditions. A delta-function mass has a dimensionless integrated strength and can preserve the conformal symmetry of the plane \cite{Bobev:2013yra}. The half-BPS M-theory Janus solution considered below realizes the $\gamma=1$ interface symmetry. At $k=1$ the supersymmetry enhancement involves additional charges beyond the manifest ABJM construction. We will test the interface interpretation using its flux and the M2 probe.


\subsection{Supergravity solution in M-theory}


We recall the solutions of \cite{DHoker:2008lup,DHoker:2008rje,Bachas:2013vza}, in the conventions of \cite{Bena:2025hxt}, since they are used both here and in the last section of this paper. The metric takes the form of a warped product of $\AdS_3\times S^3\times S^3$ over a Riemann surface $\Sigma$ with boundary: 
\begin{equation}
    ds^2_{11}=e^{2A}\Big(f_1^2\,ds^2_{\AdS_3}+f_u^2\,ds^2_{S^3_u}+f_v^2\,ds^2_{S^3_v}+\frac{|\partial_\zeta h|^2}{h^2}|d\zeta|^2\Big)\,,
\end{equation}

where $ds^2_{\AdS_3}$ is the metric on the unit radius $\AdS_3$, and $ds^2_{S^3_{u/v}}$ the metric on a unit radius three sphere. We choose complex coordinates on the upper half plane $\zeta=\xi+i\rho$ following \cite{Bachas:2013vza}. The function $h$ is harmonic on $\Sigma$ and the complex function $G$ obeys:
\begin{equation}
    \partial_\zeta G=\frac12\,(G+\bar G)\,\partial_\zeta\ln h\,.
\label{eq:Geq}
\end{equation}
We also introduce:
\begin{equation}
    W_\pm=|G\pm i|^2+\gamma^{\pm1}\big(|G|^2-1\big)\,,
\end{equation}
so that the metric functions are simply given by:
\begin{align}
 f_u^{-2}=W_+\,,&\qquad f_v^{-2}=W_-\,,\\
f_1^{-2}=\frac{(1+\gamma)^2}{\gamma}\big(|G|^2-1\big)\,,&\quad e^{6A}={\rm sign}(\gamma)\,h^2\big(|G|^2-1\big)W_+W_-\,.
\label{eq:BDEKfun}
\end{align}
The equations~\eqref{eq:Geq} for $(h,G)$ do not involve $\gamma$: the parameter only enters through the map~\eqref{eq:BDEKfun} to eleven-dimensional fields. Regularity requires :
\begin{equation}
    {\rm sign}(\gamma)(|G|^2-1)>0\,.
\end{equation}
The three-form is :
\begin{equation}
    C_3=b_1\text{Vol}(\AdS_3)+b_u\text{Vol}(S^3_u)+b_v\text{Vol}(S^3_v)
\end{equation}
With, 
\begin{align}
b_1&=\frac{\nu_1}{\sigma_1^3}\Big(\frac{h(G+\bar G)}{|G|^2-1}+\frac{(1+\gamma)^2}{\gamma}\,\Phi-\big(\gamma-\gamma^{-1}\big)\tilde h\Big)\,,\\
b_u&=\frac{\nu_2}{\sigma_2^3}\Big(-\gamma\,\frac{h(G+\bar G)}{W_+}+\gamma\big(\Phi-\tilde h\big)\Big)\,,\qquad
b_v=\frac{\nu_3}{\sigma_3^3}\Big(\frac1\gamma\,\frac{h(G+\bar G)}{W_-}-\frac1\gamma\big(\Phi+\tilde h\big)\Big)\,,
\label{eq:bfluxes}
\end{align}
where $\Phi$ is the potential we shall shortly introduce and:
\begin{align}
    &\sigma_1={\rm sign}(\gamma)|\gamma|^{1/2}+|\gamma|^{-1/2}\\
    &\sigma_2=-{\rm sign}(\gamma)|\gamma|^{1/2}\\
    &\sigma_3=-|\gamma|^{-1/2}\,,
\end{align}
are the structure constants of $D(2,1|\gamma)$ normalized so that $\sigma_1+\sigma_2+\sigma_3=0$ and $\gamma=\sigma_2/\sigma_3$, and $\nu_i=\pm1$ are signs with $\nu_1\nu_2\nu_3=-{\rm sign}(\sigma_1\sigma_2\sigma_3)$. We take $(\nu_1,\nu_2,\nu_3)=(1,1,-1)$ for $\gamma>0$ and $(-1,-1,1)$ for $-1<\gamma<0$. The solutions with $\gamma<-1$ are obtained by exchanging $\mathbb{R}^4_u$ and $\mathbb{R}^4_v$. And the real potentials are defined by
\begin{equation}
 \partial_\zeta\Phi=\overline G\,\partial_\zeta h,\qquad
 \rho\partial_\xi\widetilde\Phi=-\partial_\rho\Phi,\qquad
 \rho\partial_\rho\widetilde\Phi=\partial_\xi\Phi\,.
 \label{eq:Phidef}
\end{equation}
In the range $\gamma>0$ or $-1<\gamma<0$, the map from brane coordinates to $\AdS$ ones, with the trivial harmonic function $h=2\rho$ reads\cite{Bena:2025hxt}:
\begin{align}
 u&=a_\gamma~\mu^\alpha ~\rho^{1/2}e^{-\widetilde\Phi/4}\,,&
 z&=-\frac{1+\gamma}{4\gamma}~\mu^{-2\alpha}~\rho^{-1}e^{\widetilde\Phi/2}(\Phi+2\xi)\,,\nonumber\\
 v&=a_\gamma~\mu^{1-\alpha}~\rho^{1/2}e^{\widetilde\Phi/4}\,,&
 w&=\frac{1+\gamma}{4}~\mu^{-2(1-\alpha)}~\rho^{-1}e^{-\widetilde\Phi/2}(\Phi-2\xi)\,,\label{eq:Mmapgeneral}\\
 \alpha&=\frac\gamma{1+\gamma}\,,& a_\gamma^2&=\frac{2\sqrt{|\gamma|}}{1+\gamma}\,.\nonumber
\end{align}
Additive constants in the potentials fix the origin and gauge of the brane variables and must be chosen consistently when comparing probes. Note that the map is local, and it is always possible to choose complex coordinates such that $h$ is trivial, hence the mapping can be most generally written as:
\begin{align}
 u&=a_\gamma~\mu^\alpha ~\left(\frac{h}{2}\right)^{1/2}e^{-\widetilde\Phi/4}\,,&
 z&=-\frac{1+\gamma}{4\gamma}~\mu^{-2\alpha}~\left(\frac{h}{2}\right)^{-1}e^{\widetilde\Phi/2}(\Phi-\tilde{h})\,,\nonumber\\
 v&=a_\gamma~\mu^{1-\alpha}~\left(\frac{h}{2}\right)^{1/2}e^{\widetilde\Phi/4}\,,&
 w&=\frac{1+\gamma}{4}~\mu^{-2(1-\alpha)}~\left(\frac{h}{2}\right)^{-1}e^{-\widetilde\Phi/2}(\Phi+\tilde{h})\,,\label{eq:Mmapgeneral}\\
 \alpha&=\frac\gamma{1+\gamma}\,,& a_\gamma^2&=\frac{2\sqrt{|\gamma|}}{1+\gamma}\,.\nonumber
\end{align}
It is immediate to see that the metric is invariant under the scaling symmetry responsible of the conformal invariance of the solution:
\begin{equation}
 (u,v,z,w,x^0,x^1) \rightarrow
 (\Lambda^\alpha u,\Lambda^{1-\alpha}v,
 \Lambda^{-2\alpha}z,\Lambda^{-2(1-\alpha)}w,\Lambda x^0,\Lambda x^1)\,.
 \label{eq:scaling}
\end{equation}
The two transverse $\mathbb{R}^4$ have the same weight only at $\gamma=1$, for the Janus solution in M-theory, we will restrict ourselves to $\gamma=1$.


\subsection{The Janus solution and its expansion}
\label{sec:JanusExpM}

We use strip coordinates $(x,y)$ to reproduce the analysis of the previous section, such that $x\in ]-\infty,+\infty[$ and $y\in [0,\frac{\pi}{2}]$. With $h$ required to vanish at the two boundaries of the strip, the solution is given by:
\begin{equation}
 h=8h_0\cosh(2x)\sin(2y)\,,\qquad
 G=i\frac{\cosh(2x)+i\lambda\sin(2y)}{\cosh(2x-2iy)}\,,
 \qquad\gamma=1\,.
 \label{eq:MJanusdata}
\end{equation}
The two ends are asymptotic $\AdS_4\times S^7$ regions. The upper-half-plane coordinate is $\zeta=4h_0\sinh(2x+2iy)$, and:
\begin{align}
 \Phi&=8h_0[-\sinh(2x)+\lambda\cos(2y)]\,,\nonumber\\
 \widetilde\Phi&=2[-\log\tan y+\lambda\arctan\sinh(2x)]\,,
 \qquad\widetilde h=-8h_0\sinh(2x)\cos(2y)+C_0\,,
 \label{eq:MJanuspot}
\end{align}
where $C_0$ is an arbitrary constant. These are coordinates which makes explicit the flat D-brane picture of the supergravity solutions\ref{sec:Janus interface in type IIB}.

We restrict ourselves to $\gamma=1$, $\alpha=1/2$ and $a=1$, a more complete analysis at $\gamma\ne1$ can be found in \cite{Bobev:2013yra} using different techniques. We choose the additive gauge constant $C_0$ so that the asymptotic relation between $w$ and $z$ has no $1/u^2$ or $1/v^2$ remainder, so that $C_0=8 h_0\lambda$. The map becomes:
\begin{align}
 \tan y&=\frac uv e^{\lambda\arctan\sinh(2x)}\,,&
 uv&=4h_0\mu\cosh(2x)\sin(2y),\nonumber\\
 u^2z&=8h_0\sin^2y\left(\sinh(2x)+\lambda)\right)\,,&
 v^2w&=-8h_0\cos^2y\left(\sinh(2x)-\lambda\right)\,.
 \label{eq:MJanusmap}
\end{align}
Eliminating $y$ gives a single equation for the last strip coordinate $x=x(u,v,z,\lambda)$:
\begin{equation}
 \sinh(2x)+\lambda=\frac{z}{8h_0}
 \left(u^2+v^2e^{-2\lambda\arctan\left(\sinh(2x)\right)}\right)\,.
 \label{eq:Mimplicit}
\end{equation}
We thus obtain the function entering the brane metric as follows:
\begin{equation}
 w(u,v,z)=-8h_0\,
 \frac{\sinh(2x)-\lambda}
 {v^2+u^2e^{2\lambda\arctan\left(\sinh(2x)\right)}}\,.
 \label{eq:Mexactw}
\end{equation}
We should choose the branch continued from $\sinh(2x)=z(u^2+v^2)/(8h_0)$ at $\lambda=0$. These equations determine the non perturbative $w(u,v,z)$ without introducing an additional inverse-map variable, the full map is nonetheless not explicitely invertible. 

The pre-potential for $\AdS_4\times S^7$ is obtained by taking the $\lambda\rightarrow0$ limit of the change of coordinates of \eqref{eq:Mexactw} which leads to $w=-z$, and then solve the Monge-Ampère equation, which leads to:
\begin{equation}
 G^{(0)}=-\frac{z^2}{2}-\frac{32h_0^2}{u^2(u^2+v^2)}\,,\qquad
 w=-z\,,\qquad H=e^{-3A_0}=\frac{L^6}{(u^2+v^2)^3}\,,\qquad L^6=256h_0^2\,.
 \label{eq:Mvac}
\end{equation}
Here $L$ is the $S^7$ radius and the $\AdS_4$ radius is $L/2$:
\begin{equation}
 L^6=32\pi^2N\lp^6\,.
\end{equation}
The tension and electric coupling of an M2 along $(x^0,x^1,z)$ cancel at fixed $\vec u,\vec v$. The flat-brane supersymmetry projector survives in these frames. The induced metric need not be flat, and the eight transverse coordinates need not have constant kinetic terms. Both facts will be important in the last section of this paper.
Expand \eqref{eq:Mimplicit} and substitute in \eqref{eq:Mexactw}. Integrating in $z$ and fixing the transverse part with the master equation gives $G=G^{(0)}~+~\lambda ~~G^{(1)}~+~\lambda^2~G^{(2)}~+~o(\lambda^2)$, with:
\begin{align}
 G^{(1)}&=\frac{8h_0z}{r^2}+\left(z^2+\frac{64h_0^2}{r^4}\right)\arctan\left(\frac{zr^2}{8h_0}\right)\,,\nonumber\\
 G^{(2)}&=-\left(z^2-\frac{64h_0^2(v^2-u^2)}{r^6}\right)\arctan^2\left(\frac{zr^2}{8h_0}\right)
\nonumber\\
 &\hspace{1em}-\frac{16h_0z}{r^2}\arctan\left(\frac{zr^2}{8h_0}\right)-\frac{64h_0^2}{r^4}
 -\frac{16h_0^2(v^2-u^2)}{u^2v^2r^2}\,.
 \label{eq:Mseries}
\end{align}
As in IIB, the first order is odd in $z$ and angle-independent in the pre-potential. Two useful quantities are:
\begin{align}
 e^{-3A_0}&=H\left(1+\lambda\frac{v^2-u^2}{r^2}
 \left(3\arctan\left(\frac{zr^2}{8h_0}\right)-\frac{8h_0zr^2}{64h_0^2+z^2r^4}\right)+O(\lambda^2)\right),\nonumber\\
 (-\partial_z w)&=1-2\lambda\left(\arctan\left(\frac{zr^2}{8h_0}\right)+\frac{8h_0zr^2}{64h_0^2+z^2r^4}\right)+O(\lambda^2)\,.
 \label{eq:Mfirstmetric}
\end{align}

The first contains the scalar harmonic associated with $\mathcal{O}_1$. The second interpolates between $e^{\mp\pi\lambda}$, with width of order $8h_0/r^2$. It is what looks like the IIB dilaton profile but is a metric anisotropy. We introduce the notation equivalent to that of \ref{sec:PolchinskiBrief} but in M-theory in appendix \ref{app:BenaBrief}.

The three form gauge fields follow:
\begin{equation}
 C_3^{(1)}=\frac{2048h_0^3\lambda}{r^4(64h_0^2+r^4z^2)}S_3\,.
 \label{eq:Mlocalized}
\end{equation}
The harmonic is the $35_c$ mass harmonic. With $\nu=r^2/(8h_0)$ at zero deformation, its profile is the integral of the dimension-two kernel along the interface:
\begin{equation}
 \int_{\R^2}\dd^2x_\parallel
 \frac{\nu^{-2}}{(\nu^{-2}+\tilde{z}^2+|x_\parallel|^2)^2}
 =\frac{\pi}{1+\nu^2\tilde{z}^2}\,.
\end{equation}

This reproduces the localized-source interpretation of \cite{DHoker:2009lky}, now in the variables needed for a direct comparison with Bena's perturbation.

Smearing weak interfaces gives:
\begin{equation}
 \int_{-\infty}^{\infty}\dd z\,C_3^{(1)}
 =\pi\lambda \left( \frac{256h_0^2}{r^6}\right) S_3\,.
\end{equation}
The pre-potential can be written:
\begin{equation}
 G_{\text{smear}}=G^{(0)}+\frac{\pi\lambda}{3}
 \left(z^3+\frac{192h_0^2z}{r^4}\right)+O(\lambda^2)\,.
\end{equation}
The accompanying metric perturbation is a diffeomorphism of the vacuum, generated by:
\begin{equation}
 \xi=\frac{\pi\lambda}{2}
 \left(z(u\partial_u-v\partial_v)-\left(z^2+\frac{64h_0^2}{r^4}\right)\partial_z\right).
\end{equation}
The electric potential changes by the same Lie derivative, up to gauge. This explains the difference from type IIB Janus: the linear dilaton there was physical, while this part of the M-theory metric perturbation is a coordinate effect. The uniform $SO(4)\times SO(4)$ mass deformation is itself supersymmetric.


\subsection{Probe M2-branes and the gluing condition}
\label{sec:JanusProbepM}

In this section, we repeat the computations of section \ref{sec:The six massless scalars}, which showed that the boson mass of the D3-brane vanishes in the Janus solution, leaving six massless transverse scalar fields. The action for a probe M2-brane is simply a Nambu-Goto-like action with a coupling to the three-form gauge potential:

\begin{equation}
    S_{\text{M2}}=-T_{\text{M2}}\int d^3x\sqrt{-\text{det}(g_{\alpha\beta})}-\int d^3x ~C_3\,.
\end{equation}

It is straightforward to show that an M2-brane at fixed $u,v$ feels no force. We shall now show that the bosons remain massless in the dual field theory. We expand the action around $(u,v)=(u_0,v_0)$, using static gauge with worldvolume coordinates $(x^a,z)$. Writing $u=u_0+\epsilon u(x^a,z)\,,\quad v=v_0+\epsilon v(x^a,z)$, we obtain:
\begin{equation}
    S_{\text{M2}}=-\frac{T_{\text{M2}}\epsilon^2}{2}\int d^2x dz \left(\eta^{ab}\partial_au\partial_b u+\frac{(\partial_zu)^2}{(-\partial_zw)}+(-\partial_zw)\eta^{ab}\partial_av\partial_b v+(\partial_zv)^2\right)\,.
\end{equation}
To properly normalize the transverse excitations, we introduce the coordinate $\tilde{z}$ through $d\tilde{z}=(-\partial_zw)^{\frac{1}{2}}dz$ and redefine the scalar fields as follows:
\begin{equation}
    U=(-\partial_zw)^{-\frac{1}{4}}u\,,\quad V=(-\partial_zw)^{+\frac{1}{4}}v\,.
\end{equation}

Expanding to second order in $\epsilon$ and first order in $\lambda$, we obtain:
\begin{equation}
    S_{\text{M2}}=-\frac{T_{\text{M2}}\epsilon^2}{2}\int d^2x d\tilde{z} \left(\eta^{ij}\partial_iU\partial_jU~+~\eta^{ij}\partial_iV\partial_j V+ \lambda \tilde{z} \left(-M^2(r_0,\tilde{z}) U^2+ M^2(r_0,\tilde{z}) V^2\right)\right)\,.
\end{equation}
where $M^2(r_0,\tilde{z})=4\frac{ (8h_0)^3 r_0^6}{\left((8h_0)^2+r_0^4\tilde{z}^2\right)^3}$. We recall that this mass term allows one to identify $m(\tilde{z})$ through $M_U^2=m^2-m',\,M_V^2=m^2+m'$.  At fixed nonzero $\tilde{z}$, this coefficient vanishes as $r_0\rightarrow\infty$, so the scalar fields remain massless away from the interface. The interface nevertheless imposes the matching condition \eqref{eq:jumpM}, which can be derived from the supersymmetric first-order equations. For this interface, the supersymmetric branch satisfies:
\begin{equation}
    \partial_{\tilde{z}}U=-m(\tilde{z})U\,,
    \qquad
    \partial_{\tilde{z}}V=m(\tilde{z})V\,,
    \qquad
    m(\tilde{z})=-\lambda\frac{(8h_0)^3r_0^2}
    {\left((8h_0)^2+r_0^4\tilde{z}^2\right)^2}\,.
\end{equation}
Integrating across an interval of width $2\theta$ centered at $\tilde{z}=0$ gives
\begin{align}
    \log\left(\frac{U(+\theta)}{U(-\theta)}\right)
    &=\lambda\left(
    \arctan\left(\frac{r_0^2\theta}{8h_0}\right)
    +\frac{(8h_0)r_0^2\theta}{(8h_0)^2+r_0^4\theta^2}
    \right)\,,\\
    \log\left(\frac{V(+\theta)}{V(-\theta)}\right)
    &=-\lambda\left(
    \arctan\left(\frac{r_0^2\theta}{8h_0}\right)
    +\frac{(8h_0)r_0^2\theta}{(8h_0)^2+r_0^4\theta^2}
    \right)\,.
\end{align}
Taking first $r_0$ to infinity at fixed $\theta>0$, and then the size of the interface to zero size, one obtains:
\begin{equation}
    U(0_+)=U(0_-)e^{-m_0}\,,\quad V(0^+)=e^{m_0}V(0^-)\,,\quad m_0=-\frac{\pi\lambda}{2}\,.
\end{equation}

which is exactly the field theory result, the bosons remain massless, but have to satisfy a gluing condition across the interface.


\section{The space dependent conformal mass}
\label{sec:conformalMassdef}


\subsection{Field theory: the conformal mass deformation}


The localized deformation of the previous section vanishes away from the interface. In this section, we first show that there exists a non-uniform mass profile that preserves the conformal symmetry of a plane, in addition to the “localized” Janus deformation. We then show that these mass deformations are dual to a large class of solutions found in \cite{DHoker:2008lup,DHoker:2008rje,Bachas:2013vza}. We identify the parameter $\gamma$ in the supergravity solutions with a parameter of the mass deformation.

\subsubsection*{The conformal mass deformation}

In \cite{Kim:2018qle,Kim:2019kns}, it was shown that there exist supersymmetric mass deformations with an arbitrary mass profile $m(\tilde{z})$. We now show that certain profiles can also preserve the conformal symmetry of a plane, $SO(2,2)$\cite{Herzog:2019bom}. The supersymmetric operator $\mathcal O_2$ has dimension two in the undeformed three-dimensional theory. Its source $m$ consequently has mass dimension one. Under a dilatation $D$, the fixed background source must satisfy:
\begin{equation}
 m(D\tilde{z})=\frac{1}{D}m(\tilde{z})\,.
 \label{eq:Jscaling}
\end{equation}
For a smooth profile on $\tilde{z}>0$, this condition immediately gives:
\begin{equation}
 m(\tilde{z})=\frac{c}{\tilde{z}}\,.
 \label{eq:FTconformalmass}
\end{equation}
The dimensionless parameter $c$ is defined by the fermion mass coefficient in \eqref{eq:Lm}. To prove invariance under the full conformal group preserving the plane, we must also consider special conformal transformations with $b^{\tilde{z}}=0$. Although the source depends only on $\tilde{z}$, these transformations rescale the normal coordinate by a factor that also depends on the coordinates along the interface. The profile $m(\tilde{z})=c/\tilde{z}$ transforms with the required weight one:
\begin{equation}
    m(\Omega^{\perp}(x^0,x^1,\tilde{z}))=m\left(x^0,x^1,\frac{\tilde{z}}{1-2x_ib^i+b^2(x_ix^i+\tilde{z}^2)}\right)=\left(1-2x_ib^i+b^2(x_ix^i+\tilde{z}^2)\right)m(\tilde{z})\,.
\end{equation}

which is the transformation expected for a source of mass dimension one. The scalar coefficients $m^2$ and $m'=-c/\tilde{z}^2$ have weight two, as required for their dimension-one scalar bilinears. Thus, this mass profile preserves the conformal symmetry of a plane, and its holographic dual is expected to contain an $\AdS_3$ factor. This imposes the conformal restriction on the arbitrary-profile supersymmetric construction of \cite{Kim:2018qle,Kim:2019kns} and is similar in many respects to the conformal masses introduced in \cite{Herzog:2019bom}. The bulk Poincaré supersymmetries are enhanced to a superconformal algebra by acting with plane-preserving conformal transformations, which generate their conformal partners.

Substitution into \eqref{eq:Lm} gives the Lagrangian:
\begin{equation}
 \mathcal L=\mathcal L_{\rm ABJM}-\frac{c}{\tilde{z}}\mathcal O_2
 -\frac{c^2}{\tilde{z}^2}\tr(Y_A^\dagger Y^A)
 +\frac{c}{\tilde{z}^2}\mathcal O_1\,.
 \label{eq:FTconformalaction}
\end{equation}

The fermion mass scales as inverse distance, while its scalar completion scales as inverse distance squared. Using \eqref{eq:FTquadraticgeneral}, the quadratic coefficients (the squared masses) are:

\begin{equation}
 M_U^2(\tilde{z})=\frac{c(c+1)}{\tilde{z}^2},\qquad
 M_V^2(\tilde{z})=\frac{c(c-1)}{\tilde{z}^2},.
 \label{eq:FTconformalscalars}
\end{equation}

The $c^2$ terms are coming from the constant mass deformation. The terms linear in $c$ come entirely from the extra interaction required by supersymmetry. The normal-direction operators still factorize as in \eqref{eq:susyqm}, and their commuting supersymmetric solutions are:
\begin{equation}
 U(\tilde{z})=\tilde{z}^{-c}U_0,\qquad V(\tilde{z})=\tilde{z}^cV_0\,,.
\end{equation}


The main result of the paper will be to identify $c$ and $\gamma$ as:
\begin{equation}
 c=\frac{1-\gamma}{2(1+\gamma)}\,,\qquad
 \gamma=\frac{1-2c}{1+2c}\,.
 \label{eq:FTgammadictionary}
\end{equation}
 Exchanging the two $\mathbb{R}^4$ reverses the sign of $c$. The negative $\gamma$ branch with $-1<\gamma<0$ used below corresponds to $c>1/2$. Of course such a field theory can only exist on a half line $\tilde{z}>0$.

\subsubsection*{\texorpdfstring{Field theory derivation of the super-Algebra in the $U(1)\times U(1)$ sector}{Field theory derivation of the super-Algebra in the U(1) x U(1) sector}}
\label{sec:Superalgebra}

In the previous subsection we showed that the profile $m(\tilde z)=c/\tilde{z}$ preserves the conformal group $SO(2,2)$ of the plane. We now argue that the eight Poincar\'e supercharges are completed by eight conformal ones into $D(2,1|\gamma)\oplus D(2,1|\gamma)$, and that $\gamma$ is fixed by $c$ as announced in \eqref{eq:FTgammadictionary}. 

At $k=1$, the $U(N)\times U(N)$ theory contains a decoupled free sector describing the center of mass of the M2-branes \cite{Aharony:2008ug,Bashkirov:2010kz}. In this subsector, the quartic part of $\mathcal{O}_2$ vanishes and the deformation to ABJM is quadratic (and therefore free). Its structure constants can be read off in this sector (even though we don't prove in full generality that the supergroup is $D(2,1|\gamma)$). The advantage is that $SO(4)\times SO(4)$ and the eight Poincar\'e supercharges are all manifest besides the six explicit ones in the $\mathcal{N}=6$ formulation.

In this sector, we will make the $SO(8)$ R-symmetry of the $k=1$ theory manifest and use $X^I$ to be the eight real components of the four $Y^A$, so that $X^I$ transform in the $\mathbf{8}_v$ (with the natural splitting $(U^I,V^I)$. The four complex fermions $\psi_A$ become eight three-dimensional Majorana fermions $\psi^{\dot{p}}$ in the $\mathbf{8}_c$, and finally we call $\epsilon^p$ the supersymmetry parameters which transforms in the $\mathbf{8}_s$. They combine the twelve real parameters of $\omega_{AB}$ of the $\mathcal{N}=6$ formulation, with the parameters coming from the enhancement of supersymmetry at $k=1$ \cite{Bashkirov:2010kz}. We use $\gamma^{\mu}$ for the real three-dimensional gamma matrices with $\gamma^{01\tilde{z}}=1$ and use the usual notation $\bar{\psi}=\psi^T\gamma^0$. We call $\Gamma^I$ the $SO(8)$ gamma matrices: $\{\Gamma^I,\Gamma^J\}=2\delta^{IJ}$.

It is conveninent to introduce $\Gamma_\star\equiv\Gamma^{1234}$ which splits the two $\mathbb{R}^4$s. So that :
\begin{equation}
    \Gamma_\star \Gamma_U^I=- \Gamma_U^I\Gamma_\star\,,\quad  \Gamma_\star \Gamma_V^I=+ \Gamma_V^I\Gamma_\star\,.
\end{equation}
Where we introduced the label $\Gamma_{U/V}$ to denote all the matrices with an $I$ coming from the u-plane or v-plane. The free Lagrangian become: 
\begin{equation}
\mathcal{L}=-\frac12\,\partial_\mu X^I\partial^\mu X^I+\frac{i}{2}\,\bar\psi\gamma^\mu\partial_\mu\psi-\frac{1}{2}\left(m^2-\,m'\right)X_U^IX_U^I-\frac{1}{2}\left(m^2+\,m'\right)X_V^IX_V^I+\frac{i}{2}\,m\,\,\bar{\psi}\Gamma_\star\psi\,.
\label{eq:Lfree}
\end{equation}
The supersymmetry transformations are:

\begin{align}
\delta X^I&=i\,\bar\epsilon^p\,\Gamma^I_{p\dot p}\,\psi^{\dot p}\,,\\
\delta\psi^{\dot p}&=-\Gamma^I_{p\dot p}\Big(\gamma^\mu\epsilon^p\,\partial_\mu X^I+X_U^I\big(\eta^p+\,m\,(\Gamma_\star)_{pq}\,\epsilon^q\big)+X_V^I\big(\eta^p-\,m\,(\Gamma_\star)_{pq}\,\epsilon^q\big)\Big)\,,\qquad
\epsilon^p=\epsilon_0^p+x^\mu\gamma_\mu\eta^p\,,\nonumber
\label{eq:susyfree}
\end{align}

with constant $\epsilon_0$ (Poincar\'e) and $\eta$ (conformal). For $m=0$ these are the $16+16$ superconformal transformations of the free $\mathcal{N}=8$ theory. From now on we suppress the $SO(8)$ spinor indices. For a general profile $m(\tilde z)$, a short computation shows that the variation of (\ref{eq:Lfree}) is a total derivative provided:
\begin{equation}
m'(\tilde z)\left(1-\gamma^{\tilde z}\Gamma_\star\right)\epsilon(x)=2\,m(\tilde z)\,\Gamma_\star\eta\,.
\label{eq:susycond}
\end{equation}
For $\eta=0$ this is the projection
\begin{equation}
\gamma^{\tilde z}\Gamma_\star\,\epsilon_0=\epsilon_0\,,
\label{eq:projQ}
\end{equation}
which is the $\mathcal{N}=8$ version of \eqref{eq:FTprojection}: eight Poincar\'e supercharges survive for any $m(\tilde z)$, as in \cite{Kim:2018qle,Kim:2019kns}. Since $\gamma^{\tilde z}$ is the two-dimensional chirality, the Poincar\'e supercharges of chirality $\pm1$ have $\Gamma_\star=\pm1$. For the conformal supercharges, the terms linear in $x^0,x^1$ in (\ref{eq:susycond}) vanish only if:
\begin{equation}
\gamma^{\tilde z}\Gamma_\star\,\eta=-\eta\,.
\label{eq:projS}
\end{equation}
Then $\gamma_{\tilde z}\eta=-\Gamma_\star\eta$, and \eqref{eq:susycond} reduces to
\begin{equation}
\tilde z\,m'(\tilde z)=-m(\tilde z)\,.
\end{equation}
Eight conformal supercharges are thus preserved if and only if $m=c/\tilde z$. Therefore to preserve the conformal invariance and supersymmetry, the mass deformation should be given by: $m=\frac{c}{\tilde{z}}$.

We can now compute the algebra. Using \eqref{eq:projQ} and \eqref{eq:projS}, the commutator of a Poincar\'e supersymmetry transformation and a conformal supersymmetry transformation acts on the scalars as
\begin{equation}
[\delta_{\epsilon_0},\delta_\eta],X^I=\xi^\mu\partial_\mu X^I+i\,\bar\epsilon_0\eta,X^I+i\left(1-2c\right)\left(\bar\epsilon_0\Gamma_U^{IJ}\eta\right)X_U^J+i\left(1+2c\right)\left(\bar\epsilon_0\Gamma_V^{IJ}\eta\right)X_V^J\,,
\label{eq:QScomm}
\end{equation}
where the labels $U$ and $V$ denote $SO(8)$ indices associated with the transverse scalars in the $u$- and $v$-planes, respectively, and
\begin{equation}
\xi^\mu=2i\,\bar\epsilon_0\eta\,x^\mu+2i\left(\bar\epsilon_0\gamma^\mu{}_\nu\,\eta\right)x^\nu\,.
\end{equation}
The first two terms in \eqref{eq:QScomm} describe a dilatation and a boost in the $(x^0,x^1)$ plane acting on a field of dimension $\frac{1}{2}$. They are independent of $c$ and belong to the $SO(2,2)$ factor of the bosonic symmetry group $SO(2,2)\times SO(4)_U\times SO(4)_V$. The mass parameter enters only the last two terms: rotations of $\mathbb{R}^4_u$ carry a coefficient $1-2c$, while rotations of $\mathbb{R}^4_v$ carry a coefficient $1+2c$. Comparing their relative coefficients with the conventions of \cite{Bachas:2013vza} suggests the identification:
\begin{equation}
    \gamma=\frac{1-2c}{1+2c},.
    \label{eq:gammacLink}
\end{equation}
Since this calculation is restricted to the free subsector, it does not establish the full superconformal algebra of the interacting theory. It nevertheless provides evidence for the relation between $\gamma$ and $c$ \eqref{eq:gammacLink} that we will further investigate in this section.


\subsubsection*{Field theory vacua}
\label{sec:conformalvacua}


The vacuum equations can be understood by the same method used by Kim and collaborators for inhomogeneous ABJM: rescale the scalars and change the normal coordinate to remove the position dependence of the mass \cite{Kim:2018qle,Kim:2019kns}. For $m=\frac{c}{\tilde{z}}$ this change of variables is particularly simple. It reduces conformal vacuum profiles to the classical matrix equations found in the context of mass deformed ABJM \cite{Gomis:2008vc}, whose solutions are organized by $SU(2)$ representations similar to the vacuum structure of $\mathcal{N}=1^{*}$ super Yang-Mills. The following discussion follows closely the notations of \cite{Kim:2018qle,Kim:2019kns}, in particular we use $SU(4)$ notations for the scalar fields, that we decompose into the $SU(2)\times SU(2)$ compatible with the $SO(4)\times SO(4)$ natural decomposition of the scalar fields, the $SU(2)$s are labeled by latin indices, mainly $a,b$.

A supersymmetric configuration is independent of time and of the coordinate along the plane, but the scalars are expected to depend on $\tilde{z}$. We take vanishing fermions and field strengths and choose the gauge $A_{\tilde{z}}=\widehat A_{\tilde{z}}=0$. To write the BPS equations compactly, we use the cubic term in the ABJM supersymmetry variation following \cite{Kim:2019kns}:
\begin{equation}
 \beta^{AB}{}_{C}=2\pi
 (Y^AY_C^\dagger Y^B-Y^BY_C^\dagger Y^A)\,.
\end{equation}
For $Y^a=V^a$ and $Y^i=U^{i-2}$, with $a,b=1,2$ and $i,j=3,4$, the equations are:
\begin{align}
 \partial_{\tilde{z}} Y^a-\beta^{ba}{}_{b}-\frac{c}{\tilde{z}}Y^a&=0\,,
 &\partial_{\tilde{z}} Y^i-\beta^{ji}{}_{j}+\frac{c}{\tilde{z}}Y^i&=0\,,\nonumber\\
 \beta^{ja}{}_{i}=\beta^{bi}{}_{a}
 =\beta^{ab}{}_{i}=\beta^{ij}{}_{a}&=0\,.
 \label{eq:vacfullBPS}
\end{align}
The second line constrains how the two $\mathbb{R}^4$ transverse scalars occupy the matrix space. It is satisfied by putting them on separate blocks.

Following the scalar-rescaling method of \cite{Kim:2019kns}, we write:
\begin{equation}
 s=\log(\tilde{z}/\ell)\,,\qquad
 U=\tilde{z}^{-1/2}\widetilde U(s)\,,\qquad
 V=\tilde{z}^{-1/2}\widetilde V(s)\,.
 \label{eq:vacrescaling}
\end{equation}
The reference length $\ell$ only fixes the origin of $s$. Because the interaction is cubic, both it and the derivative term carry a common factor $\tilde{z}^{-3/2}$. Cancelling this factor gives:
\begin{align}
 \frac{d\widetilde U^a}{ds}
 +(c-\tfrac12)\widetilde U^a
 -2\pi\sum_b
 (\widetilde U^b\widetilde U_b^\dagger\widetilde U^a
 -\widetilde U^a\widetilde U_b^\dagger\widetilde U^b)&=0\,,
 \nonumber\\
 \frac{d\widetilde V^a}{ds}
 -(c+\tfrac12)\widetilde V^a
 -2\pi\sum_b
 (\widetilde V^b\widetilde V_b^\dagger\widetilde V^a
 -\widetilde V^a\widetilde V_b^\dagger\widetilde V^b)&=0\,.
 \label{eq:vacautonomous}
\end{align}
In these coordinates, a dilatation is just a translation of $s$,  hence a conformal vacuum has constant $\widetilde U$ and $\widetilde V$. The derivative of the rescaling factor has shifted the two coefficients to $c-1/2$ and $c+1/2$. This is the only change needed in the matrix algebra.

Let $n$ be a positive integer and $G^1_{(n)},G^2_{(n)}$ be the $n\times(n+1)$ GRVV matrices \cite{Gomis:2008vc} (that we follow closely from now on):
\begin{equation}
 (G^1_{(n)})_{ij}=\sqrt{n-i+1}\,\delta_{ij}\,,\qquad
 (G^2_{(n)})_{ij}=\sqrt{i}\,\delta_{j,i+1}\,.
 \label{eq:vacGRVV}
\end{equation}
They obey:
\begin{equation}
 \sum_b(G^b G_b^\dagger G^a
 -G^a G_b^\dagger G^b)=G^a\,,
 \label{eq:vacGRVVidentity}
\end{equation}
while their Hermitian conjugates obey the equation with the opposite sign. The two conformal profiles are therefore:
\begin{align}
 U^a&=\sqrt{\frac{c-1/2}{2\pi\tilde{z}}}\,G^a_{(n)}\,,
 &V^a=0\,,\\
 V^a&=\sqrt{\frac{c+1/2}{2\pi\tilde{z}}}\,(G^a_{(n)})^\dagger\,,
 &U^a=0\,.
 \label{eq:vacblocks}
\end{align}

After these positive normalization factors are removed, the algebraic equations are exactly the GRVV vacuum equations. We can thus use their block classification. We have just shown that the structure of the vacua remains the same as in mABJM.

Following GRVV, the $su(2)$ vacuum structure a la Polchinski-Strassler can be made explicit. We can define $n\times n$ matrices  through:

\begin{equation}
 J_+=G^2G_1^\dagger\,,\qquad J_-=G^1G_2^\dagger\,,
 \qquad J_3=\tfrac12(G^1G_1^\dagger-G^2G_2^\dagger)\,.
\end{equation}
These matrices obey the usual $su(2)$ algebra:
\begin{equation}
 [J_3,J_\pm]=\pm J_\pm\,,\qquad [J_+,J_-]=2J_3\,,\qquad
 J_3^2+\tfrac12(J_+J_-+J_-J_+)=\frac{n^2-1}{4}\mathbf1_n\,.
 \label{eq:vacsu2}
\end{equation}
This is the spin $\frac{n-1}{2}$ representation of $su(2)$. The bilinears on the other gauge space give spin $\frac{n}{2}$, of dimension $n+1$. This rectangular structure replaces the adjoint matrices of the Polchinski-Strassler construction from bi-fundamental ones. The bilinears describe the fuzzy two-sphere base of the polarized configuration, in the M-theory interpretation the fiber completes the wrapped $S^3$ \cite{Nastase:2009ny}.

A general conformal block configuration is a direct sum of the two types in \eqref{eq:vacblocks}. Let $N_n$ count the $n\times(n+1)$ $U$ blocks and $N'_n$ the $(n+1)\times n$ $V$ blocks. The $n=0$ blocks are empty columns or rows. Requiring both gauge spaces to have dimension $N$ gives:

\begin{equation}
 \sum_n\left(nN_n+(n+1)N'_n\right)=N\,,
 \qquad \sum_n\left((n+1)N_n+nN'_n\right)=N\,.
 \label{eq:vacranks}
\end{equation}
Equivalently:
\begin{equation}
 \sum_nN_n=\sum_nN'_n\,,\qquad
 \sum_n(n+\tfrac12)(N_n+N'_n)=N\,.
 \label{eq:vacoccupation}
\end{equation}
The representation sizes and multiplicities are discrete labels, just as the irreducible $SU(2)$ factors label polarized D-brane vacua. The amplitudes in \eqref{eq:vacblocks} are fixed by $c$, so there is no independent radius modulus for a nonzero conformal block.

For homogeneous mass-deformed ABJM at $k=1$, the quantum occupation rule is $N_n,\,N'_n\in\{0,1\}$ \cite{Kim:2019kns}. With this rule, the occupied sizes in each list are distinct. List them in decreasing order as $a_1>\dots>a_d\geq0$ and $b_1>\dots>b_d\geq0$. A Young diagram is obtained by drawing $d$ diagonal boxes and attaching $a_i$ boxes to the right and $b_i$ boxes below the $i$th diagonal box. The area of the Young-Tableau is:
\begin{equation}
 |Y|=\sum_{i=1}^d(a_i+b_i+1)=N\,.
 \label{eq:vacfrobenius}
\end{equation}

The classical result for our conformal profiles is the occupation data in \eqref{eq:vacoccupation}. The ordinary Young-diagram classification follows when the $k=1$ quantum occupation rule is imposed.

Finally, the representation size has a direct geometric meaning. For a $U$ block, the radius of the fuzzy 3-sphere is:
\begin{equation}
 R_U^2= \frac{2}{n}\sum_a\tr( U^a U_a^\dagger)
 =\frac{n+1}{\pi \tilde{z}}(c-\tfrac12)\,.
\end{equation}
At large $n$, this becomes:
\begin{equation}
 R_U^2\simeq -\frac{n\gamma}{\tilde{z}\pi(1+\gamma)}\,,
\end{equation}
This is the radius of the polarization of the M2-brane into M5-branes: the matrix representation fixes the size of the expanded shell of brane, and its charge contributes to the total area of the diagram. 

We thus obtained that the vacua of the theory are labelled by Young-Tableau and that they perfectly match that of the gravity dual which we will explain in the next subsection, as was shown for ABJM \cite{Kim:2010mr} with its gravity duals given by the LLM geometries \cite{Lin:2004nb}.


\subsection{\texorpdfstring{Supergravity: the $D(2,1|\gamma)^{\oplus2}$ solutions}{Supergravity: the D(2,1,gamma) + D(2,1,gamma) solutions}}


In this section we derive the solutions which will be shown to correspond to the holographic dual of the field theory we have just introduced. The equations for $h$ and $G$ do not contain $\gamma$. A solution with $|G|>1$ therefore gives a positive-$\gamma$ family, while one with $|G|<1$ gives a negative-$\gamma$ family. We keep the sphere labels of the brane map: $S^3$ is wrapped by the M5 and $S^{\prime3}$ by the M5$'$. We focus on the $\gamma<0$ solutions with an odd number of flips so that they have an asymptotic $\AdS_4\times S^7$ region.

The bubbling data on the upper half-plane are:
\begin{align}
 G&=i\sum_{j=1}^{2p+1}(-1)^{j+1}\frac{\zeta-\xi_j}{|\zeta-\xi_j|},&
 \Phi&=-2\sum_{j=1}^{2p+1}(-1)^{j+1}|\zeta-\xi_j|,\nonumber\\
 \widetilde\Phi&=-2\sum_{j=1}^{2p+1}(-1)^{j+1}\log\tan\frac{\theta_j}{2},&
 &\zeta-\xi_j=|\zeta-\xi_j|e^{i\theta_j}.
 \label{eq:oddflip}
\end{align}
Here $\xi_1<\dots<\xi_{2p+1}$, $h=2\rho$ and $\widetilde h=-2\xi$. On the boundary $G$ alternates between $+i$ and $-i$. The $u$ sphere shrinks where $G=+i$, and the $v$ sphere where $G=-i$. Fibering these spheres over intervals produces four-cycles carrying the five-brane fluxes. In the interior $|G|<1$.

There is one asymptotic region at $|\zeta|\to\infty$. It is locally $\AdS_4\times S^7$, with a half-space conformal boundary.

We choose the origin of $\xi$ so that:
\begin{equation}
 \sum_{j=1}^{2p+1}(-1)^{j+1}\xi_j=0.
 \label{eq:moments}
\end{equation}
The radius of  $\AdS_4\times S^7$  fixes the M2 charge:
\begin{equation}
 L^6=16\sum_{j=1}^{2p+1}(-1)^{j+1}\xi_j^2
 =32\pi^2N\lp^6.
 \label{eq:DN}
\end{equation}

We take $-1<\gamma<0$. On each shrinking-sphere segment the corresponding three-form potential is constant. Its jump between adjacent such segments gives the four-form flux through the cycle under consideration. From \eqref{eq:bfluxes}, the magnitudes of the jumps are $4\ell/\sqrt{-\gamma}$ for $b_u$ and $4\sqrt{-\gamma}\ell$ for $b_v$. With $(2\pi\lp)^{-3}\int F_4\in\mathbb Z$,
\begin{equation}
 \ell_{(u)}=\frac{\pi\lp^3}{\sqrt{-\gamma}}N_u\,,\qquad
 \ell_{(v)}=\pi\lp^3\sqrt{-\gamma}N_v\,.
 \label{eq:segmentcharges}
\end{equation}
The label $(u)$ denotes a segment on which $S^3$ shrinks, and $N_u$ counts the M5-branes wrapping $S^3$. Their charge is measured on the linking four-cycle built from the other sphere. It was shown in \cite{Bena:2025hxt} that the position of a new bubble with a small charge coincides with the position of a probe M2-brane that polarizes into a spike-shaped M5-brane. This parallels the computation of \cite{Massai:2014wba}, which relates a small droplet in the LLM geometries to a probe M2-brane polarizing into an M5-brane. For a more detailed discussion of these quantization conditions and their interpretation, see \cite{Bachas:2013vza,Bena:2025hxt}.

For three flips, $\ell_{(u)}=\xi_2-\xi_1$ and $\ell_{(v)}=\xi_3-\xi_2$. One finds:
\begin{equation}
 L^6=32\ell_{(u)}\ell_{(v)}\,,\qquad
 \sum_j(-1)^{j+1}\xi_j^3=3\ell_{(u)}\ell_{(v)}(\ell_{(u)}-\ell_{(v)})\,,\qquad N=N_uN_v\,.
\end{equation}
For several alternating intervals, the Young tableau is:
\begin{equation}
 Y=\left[\left(N_u^{(1)}+\cdots+N_u^{(p)}\right)^{N_v^{(p)}},
 \ldots,\left(N_u^{(1)}\right)^{N_v^{(1)}}\right].
 \label{eq:diagram}
\end{equation}
A row is an M5$'$ wrapping $S^{\prime3}$, with length equal to its attached M2 charge. A column is an M5 wrapping $S^3$. We denote row lengths by $\Lambda_i$ and column lengths by $\Lambda'_j$. The total number of boxes of $Y$ is:
\begin{equation}
 |Y|=\frac{L^6}{32\pi^2\lp^6}=N\,.
\end{equation}


\subsection{M2 probe brane and the boson mass}


By construction, the solution preserves the Killing spinors of the flat M2-branes along $(x^a,z)$, and hence a probe M2-brane at fixed $(u,v)$ is floating. This section is devoted to showing that despite the existence of these flat directions, the mass of the transverse scalar fields is non trivial in the dual field theory. This is to be contrasted with the analysis of \cite{Polchinski:2000uf,Bena:2000zb} where the solution was shown not to have Killing spinors of the underformed theory:  flat D3-branes for Type IIB and flat M2 branes for M-theory. We follow the same logic as in section \ref{sec:The six massless scalars}, and especially linearizing the probe M2-brane around a fixed $(u,v)=(u_0,v_0)$ solution leads to using coordinates $d\tilde{z}=(-\partial_zw)^{\frac{1}{2}} dz$ and scalar fields: $U=(-\partial_zw)^{-\frac{1}{4}}u\,,\quad V=(-\partial_zw)^{+\frac{1}{4}}v$. We obtain:
\begin{align}
\label{eq:M2probeBubbling}
    S_{\text{M2}}=-\frac{T_{\text{M2}}\epsilon^2}{2}\int d^2xd&\tilde{z}~\Bigg(\eta^{ij}\partial_i U\partial_jU ~+~ \eta^{ij}\partial_i V\partial_jV\\
    &+\left(\frac{\partial_z^3w}{4(\partial_zw)^2}+\frac{7(\partial_z^2w)^2}{16(-\partial_zw)^3}\right)_{u_0,v_0}U^2+\left(-\frac{\partial_z^3w}{4(\partial_zw)^2}-\frac{5(\partial_z^2w)^2}{16(-\partial_zw)^3}\right)_{u_0,v_0}V^2\Bigg)\,.\nonumber
\end{align}

The rest of the analysis is a bit more intricate than in the Janus interface situation. The supergravity solution we consider in this section only have one $\AdS_4\times S^7$ region, while an other one is a smooth closing of spacetime. Moreover it was possible to view the Janus-interfaces as perturbations of $\AdS_5\times S^5$ (resp. $\AdS_4\times S^7$) in a small parameter $\delta$ (resp. $\lambda$), this is not possible here. We shall therefore first consider an expansion of the solution "far from the sources" to identify the mass deformation. We first introduce $\zeta=R e^{i\theta}$ the radial coordinate on the Riemann surface centered at $h=0$. Then an expansion of the metric at large $R$ leads to:

\begin{equation}
    ds^2=\frac{L^2}{4}\left(\frac{dR^2}{R^2}-\gamma\frac{64 R^2}{(1+\gamma)^2L^6}ds_{\AdS_3}^2\right)+L^2\left(\frac{1}{4}d\theta^2+\cos(\frac{\theta}{2})^2ds_{S_{3v}}^2+\sin(\frac{\theta}{2})^2ds_{S_{3u}}^2\right)\,.
\end{equation}
We obtain for the coordinates in this large $R$ expansion
\begin{align}
\label{eq:MtheoryExp}
 u&=a_\gamma\sqrt{2R}\,\mu^\alpha\sin\frac{\theta}{2}
 \left(1+\frac{L^6\cos\theta}{64R^2}+O(R^{-3})\right),\\
 v&=a_\gamma\sqrt{2R}\,\mu^{1-\alpha}\cos\frac{\theta}{2}
 \left(1-\frac{L^6\cos\theta}{64R^2}+O(R^{-3})\right),\\
 z&=\frac{1+\gamma}{2\gamma}\,\mu^{-2\alpha}
 \left(1+\frac{L^6}{32R^2}+O(R^{-3})\right),\\
 w&=-\frac{1+\gamma}{2}\,\mu^{-2(1-\alpha)}
 \left(1+\frac{L^6}{32R^2}+O(R^{-3})\right)\,,
\end{align}
where:
\begin{equation}
 \alpha=\frac{\gamma}{1+\gamma}\,,
 \qquad a_\gamma^2=\frac{2\sqrt{-\gamma}}{1+\gamma}\,.
\end{equation}
It is therefore straightforward to obtain the mass of the scalar fields by computing the term which remains constant at $R\rightarrow\infty$ in the mass parameters \eqref{eq:M2probeBubbling}:
\begin{equation}
    M_U^2=\left(\frac{\partial_z^3w}{4(\partial_zw)^2}+\frac{7(\partial_z^2w)^2}{16(-\partial_zw)^3}\right)_{u_0,\,v_0}\,,\quad M_V^2=\left(-\frac{\partial_z^3w}{4(\partial_zw)^2}-\frac{5(\partial_z^2w)^2}{16(-\partial_zw)^3}\right)_{u_0,\,v_0}\,.
\end{equation}
It leads to:
\begin{equation}
    M_U^2=\frac{(1-\gamma)(\gamma+3)}{4(1+\gamma)^2(\tilde{z})^2}\,,\quad M_V^2=\frac{(\gamma-1)(3\gamma+1)}{4(1+\gamma)^2(\tilde{z})^2}\,.
\end{equation}
These masses reproduce exactly the boson mass of the field theory with the identification:
\begin{equation}
    \gamma=\frac{1-2c}{1+2c}\,.
\end{equation}

The computation we did in this subsection is only valid at $\gamma<0$. Nonetheless it is interesting to note that a similar computation at $\gamma=1$, or for the solutions of \cite{Bena:2026vlv} would lead to no mass deformation at all, as is expected from the solution being conformal boundary conditions of ABJM or of $\mathcal{N}=4$ SYM.


\subsection{An asymptotic expansion of the four form fluxes and the fermion mass}


In this subsection we perform an asymptotic expansion of the four-form flux and show that it is the one sourced by the operator added to the ABJM Lagrangian in \eqref{eq:FTconformalaction}. We recall that the bulk field dual to an operator of a given conformal dimension admits two independent solutions of the linearized supergravity equations near the boundary: the non-normalizable one is the source of the operator, while the normalizable one encodes its expectation value. We therefore extract the coefficient of the non-normalizable mode, which is identified as the coefficient of the operator in the Lagrangian: the mass dependent position $m(\tilde{z})$.

We denote the eleven-dimensional field strength by $G_4=dC_3$. In brane coordinates,
\begin{align}
 G_4~~&=~-d\left(e^{3A_0}\sqrt{-\partial_zw}\,dt\wedge dy\wedge
 \left(dz+\frac{\partial_uw}{\partial_zw}du\right)\right)+G_4^{\rm mag}\,,\\
G_4^{\rm mag}~&=~
 d\left(u^3\frac{\partial_uw}{\partial_zw}\right)\wedge\vol(S_u^3)
 +d(v^3\partial_vw)\wedge\vol(S_v^3)\,.
 \label{eq:fluxexact}
\end{align}
Let us expand the explicit expression for $w(u,v,z)$ at large $(u,v)$, we obtain, keeping the first non trivial terms at which $\partial_uw,\,\partial_vw$ are non trivial, it leads to:

\begin{equation}
w(u,v,z)=-\frac{1+\gamma}{2}\left(\frac{2\gamma z}{1+\gamma}\right)^{1/\gamma}-\frac{(1-\gamma)L^6}{4(1+\gamma)}\frac{\left(\frac{2\gamma z}{1+\gamma}\right)^{1/\gamma}}{\left(u^2\left(\frac{2\gamma z}{1+\gamma}\right)
+v^2\left(\frac{2\gamma z}{1+\gamma}\right)^{1/\gamma}\right)^2}\,.\label{eq:fluxwexpansion}
\end{equation}

The second term scales as $R^{-2}$ at fixed $(\mu,\theta)$ \eqref{eq:MtheoryExp}. Therefore in In $b_u=u^3\frac{\partial_uw}{\partial_zw}$ it is enough to use  the unperturbed $\partial_zw$ in the denominator at this order.

To compare with a constant antisymmetric tensor on the transverse eight-plane, we need the same asymptotic transverse normalization as for the M2 scalars. At general $\gamma$: $u^2+v^2$ is not the radius in the normalized asymptotic eight-plane. The sphere angles are unchanged. In particular, we recall that we defined $\tilde{z}$ through $d\tilde{z}=(-\partial_zw)^{\frac{1}{2}}dz$ so that differentiating the fluxes read:

\begin{equation}
 b_u=-\frac{(1-\gamma)L^6}{1+\gamma}
 \frac{u^4\tilde{z}^{(\gamma-3)/(1+\gamma)}}{r^6}\,,\quad
 b_v=+\frac{(1-\gamma)L^6}{1+\gamma}
 \frac{v^4\tilde{z}^{(1-3\gamma)/(1+\gamma)}}{r^6}\,.
 \label{eq:fluxbexpansion}
\end{equation}
where we introduced $r^2$ defined by:
\begin{equation}
 u^2\left(\frac{2\gamma z}{1+\gamma}\right)
 +v^2\left(\frac{2\gamma z}{1+\gamma}\right)^{1/\gamma}
 =\tilde{z} r^2\,.
\end{equation}
It is therefore natural to define the hatted quantities $\hat{u},\,\hat{v}$ so that:
\begin{equation}
 \widehat{\vec u}=\tilde{z}^{\frac{\gamma-1}{2(1+\gamma)}}\vec u\,,
 \qquad \widehat{\vec v}=\tilde{z}^{\frac{1-\gamma}{2(1+\gamma)}}\vec v\,,
 \qquad r^2=\widehat u^2+\widehat v^2\,.
 \label{eq:fluxframe}
\end{equation}
Which are the ones from which we can see the normalized asymptotic eight-plane.

Use the flat orientation $d^4\widehat u\wedge d^4\widehat v$ and define the unit harmonic:
\begin{align}
 T_4&=d^4\widehat u-d^4\widehat v\,,\qquad \star_8T_4=-T_4\,,\label{eq:MtheoryT4}\\
 S_3&=\widehat u^4\vol(S_u^3)-\widehat v^4\vol(S_v^3)\,,
 \qquad V_4=\frac{dr}{r}\wedge S_3\,.
\end{align}
These are the tensor structures used by Bena. In this normalization:
\begin{equation}
 d_8S_3=4T_4\,,\qquad d_8V_4=-4\frac{dr}{r}\wedge T_4\,,
 \qquad \star_8V_4=-T_4+V_4\,.
 \label{eq:fluxharmonics}
\end{equation}
Here $d_8$ differentiates at fixed $\tilde{z}$. The transverse mass tensor is anti-self-dual, the complete radial four-form need not itself be anti-self-dual.

The result of the coordinate expansion is:
\begin{equation}
 C_3^{\rm int}=-\frac{1-\gamma}{1+\gamma}
 \frac{L^6}{\tilde{z} r^6}S_3+\dots\,.
 \label{eq:fluxpotential}
\end{equation}

Taking the full exterior derivative, including $\tilde{z}$, gives the source contribution:

\begin{equation}
G_{4,\text{ source}}^{\text{ mag}}=
 -\frac{1-\gamma}{1+\gamma}\frac{L^6}{\tilde{z} r^6}
 (4T_4-6V_4)
 +\frac{1-\gamma}{1+\gamma}\frac{L^6}{\tilde{z}^2r^6}
 d\tilde{z}\wedge S_3\,.
 \label{eq:fluxsource}
\end{equation}

There are three different checks. First, its transverse part is exactly the $r^{-6}(4T_4-6V_4)$ non-normalizable mass mode, rather than the $r^{-8}(4T_4-8V_4)$ response mode. Second, its relative sphere sign is the $\mathbf{35}_{-}$ mass harmonic. Third, the longitudinal term is required by the Bianchi identity: the derivative of $\tilde{z}^{-1}$ multiplying the transverse mode cancels the transverse derivative of the $d\tilde{z}\wedge S_3$ term. Keeping just the transverse mode with a varying coefficient would violate $dG_4=0$.

For a constant mass, Bena's transverse equation is
$d_8(Z^{-1}(\star_8G_4^{\text{ mag}}-G_4^{\rm mag}))=0$, with $Z=L^6/r^6$.
For our leading transverse term:
\begin{equation}
 Z^{-1}\left(\star_8G_{4,\,\perp}^{\text{ mag}}-G_{4,\,\perp}^{\text{mag}}\right)
 =\frac{2(1-\gamma)}{(1+\gamma)\tilde{z}}T_4=4 m(\tilde{z})T_4\,.
 \label{eq:fluxdualcombination}
\end{equation}
It is closed under $d_8$, as required. Its full derivative is nonzero. Thus the constant-mass equation alone is not an adequate equation of motion check for a spacetime-dependent source. The mixed flux and subleading radial terms must also be kept. From these expressions and with our conventions (summarized in \ref{app:BenaBrief}) the mass is  identified as:
\begin{equation}
    m(\tilde{z})=\frac{(1-\gamma)}{2(1+\gamma) \tilde{z}}\,,\quad c=\frac{(1-\gamma)}{2(1+\gamma)}\,,
\end{equation}
which is consistent with the field theory result.


\subsection{Open questions and outlook}


An important question concerns the claim of \cite{Bachas:2013vza} that the parameter $\gamma$ of $D(2,1|\gamma)$ is rational because it is a ratio of five-brane charges. No such constraint is apparent in our field-theory analysis, where the mass deformation parameter $c$ is continuous. It would therefore be interesting to revisit the assumptions underlying the charge-ratio argument or determine whether additional quantum or global consistency conditions restrict the allowed values of $c$. 

It would also be interesting to understand the solutions of \cite{Bachas:2013vza} that are asymptotic to $\AdS_7\times S^4$ and exhibit a similar bubbling structure. A related direction is to carry out an analogous analysis for the M-theory Janus deformations of $\AdS_7\times S^4$ constructed in \cite{Conti:2024rwd}, and to extend our framework to the quarter-BPS solutions of \cite{Arav:2018njv}.

Finally, a broader question concerns the boundary conditions of ABJM and its deformations, which must be understood to fully identify the holographic duals of these solutions. Recent work in this direction includes \cite{Okazaki:2025rjl}, which is restricted to $\gamma=1$ and the gauge group $U(1)_k\times U(1)_{-k}$. It would be interesting to investigate whether the boundary conditions studied there extend to $U(N)_k\times U(N)_{-k}$ and admit a $\gamma$-dependent generalization that could account for the full family of solutions with $\gamma>0$ \cite{Bachas:2013vza,Bena:2025hxt}.

Let us recall that the mass-deformed M2-brane theory with constant mass $m$, at inverse temperature $\beta$, is expected to have a phase transition between a black-hole phase and an horizonless bubbling saddle. The transition is controlled by $\beta m$, the only dimensionless parameter of the problem, and is the analogue of the Hawking-Page transition \cite{Witten:1998zw}, it was studied in detail for the Polchinski-Strassler theory in \cite{Freedman:2000xb,Bena:2018vtu}. Our position-dependent mass makes the finite-temperature problem particularly interesting. Since $m=\frac{c}{\tilde{z}}$, at fixed $\beta$ one has $\beta m\ll 1$ at large $\tilde{z}$ and $\beta m\gg 1$ at small $\tilde{z}$, so one may expect the dual geometry to realize a geometric Hawking-Page transition. A natural candidate is obtained by replacing $\AdS_3$ by a BTZ black hole. Since BTZ is locally $\AdS_3$, the equations of motion are still solved, in the same spirit as the $\AdS_4$ black holes of \cite{Uhlemann:2021nhu}. However, the conformal mass deformation is by construction constant in units of the $\AdS_3$ radius, and accordingly these solutions show no geometric phase transition. In fact, their horizon extends over the whole Riemann surface and reaches the conformal boundary, so they are not the flat half-space at finite temperature. We expect the true dual of the latter to have a horizon at finite depth in the asymptotic $\AdS_4\times S^7$ region, ending on the bubbling region. The black hole then localizes, and the Hawking-Page transition is realized as a spatial phase boundary.

Finally, it would also be interesting to find the $\mathcal{N}=0^{*}$ version of the conformal, position-dependent mass deformation of $\mathcal{N}=4$ SYM, whose holographic dual should be a nonsupersymmetric version of $\AdS_4\times S^2\times S^2$ fibered over a Riemann surface.

\section*{Acknowledgments}
The author would like to thank Iosif Bena, Jan de Boer, Andreas Karch, Emil Martinec and Cumrun Vafa for interesting discussions. I am grateful to Iosif Bena for his comments on an early draft of this paper.

\section*{AI statement}
Generative language models (OpenAI Astra and Claude Opus 5.5) were used for part of
the work presented in section \ref{sec:Janus interface in Type Mtheory}. They were asked to reproduce, for the
M-theory Janus solution, the analysis of section \ref{sec:Janus interface in type IIB}.
In particular, the computations of sections \ref{sec:JanusExpM}
and \ref{sec:JanusProbepM} were done by these models and then completely checked
by the author, who takes full responsibility for them. At a late stage of the project,
generative models were also used to produce referee-style reports on the manuscript.
The derivation of the superconformal algebra in the free sector of the mass deformed
theory in section \ref{sec:Superalgebra} was suggested by a model in that context, all
the computations of that section were done by the author.

\bigskip
\bigskip
\appendix
\leftline{\LARGE \bf Appendices}

\section{A lightening review of the mass deformation of ABJM in supergravity }
\label{app:BenaBrief}

This section is devoted to reproducing the computations of section \ref{sec:PolchinskiBrief}  but in M-theory. Such a deformation was explained in \cite{Bena:2000zb}, which we closely follow. We consider perturbations around $\AdS_4\times S^7$, with:
\begin{equation}
\begin{aligned}
ds_{11}^2&=Z^{-\frac{2}{3}}\eta_{\mu\nu}dx^\mu dx^\nu
+Z^{\frac{1}{3}}dx^i dx^i\,,\\
C_3^{(0)}&=-Z^{-1}dx^0\wedge dx^1\wedge dx^2\,,
\qquad Z=\frac{R^6}{r^6}\,,
\qquad R^6=32\pi^2N\ell_p^6\,.
\end{aligned}
\end{equation}
Here $\mu=0,1,2$ and $i=3,\ldots,10$. For a transverse perturbation $F_4^{(1)}$, the linearized equations are:
\begin{equation}
{\rm d}G_4^{(1)}=0\,,\qquad
{\rm d}\left[Z^{-1}\left(*_8G_4^{(1)}-G_4^{(1)}\right)\right]=0\,,
\end{equation}
where $*_8$ acts on the flat transverse eight-plane.

The traceless fermion mass matrix transforms in the $\mathbf{35}_{-}$, represented by a constant anti-self-dual tensor. Following Bena, we define the unit harmonic tensor like that of \eqref{eq:MtheoryT4}:
\begin{equation}
\begin{aligned}
T_4&=\frac{1}{4!}T_{ijkl}\,dx^i\wedge dx^j\wedge dx^k\wedge dx^l\,,
\qquad *_8T_4=-T_4\,,\\
S_3&=\frac{1}{3!}T_{ijkl}x^i\,dx^j\wedge dx^k\wedge dx^l\,,
\qquad V_4=\frac{{\rm d}r}{r}\wedge S_3\,.
\end{aligned}
\end{equation}
These satisfy:
\begin{equation}
\begin{aligned}
{\rm d}S_3&=4T_4\,,\qquad
{\rm d}V_4=-4\frac{{\rm d}r}{r}\wedge T_4\,,\\
*_8V_4&=-T_4+V_4\,,\qquad
{\rm d}(r^pS_3)=r^p(4T_4+pV_4)\,.
\end{aligned}
\end{equation}
The equations admit the two radial solutions:
\begin{equation}
G_{4,\mathrm{source}}^{(1)}
=\alpha\left(\frac{R}{r}\right)^6(4T_4-6V_4)\,,
\qquad
G_{4,\mathrm{response}}^{(1)}
=\beta\left(\frac{R}{r}\right)^8(4T_4-8V_4)\,.
\end{equation}
The first is the non-normalizable fermion mass mode, the second describes the corresponding expectation value. The source potential is:
\begin{equation}
C_3^{(1)}=\alpha\left(\frac{R}{r}\right)^6S_3\,.
\end{equation}
The mass can be identified from this formula, in particular:
\begin{equation}
    \alpha=-2m\,.
\end{equation}
In particular it leads to :
\begin{equation}
    \frac{1}{Z}\left(\star_8G_4-G_4\right)=4m T_4\,.
\end{equation}

\bibliographystyle{utphys}      

\bibliography{references}

\end{document}